\documentclass{aa}  

\usepackage{graphicx}
\usepackage{lscape}
\usepackage{txfonts}
\usepackage{color}
\begin{document} 

\title{ALMA observations of the young protostellar system Barnard 1b: signatures of an incipient hot corino in B1b-S}

\author{
N. Marcelino \inst{1}
\and
M. Gerin \inst{2}
\and
J. Cernicharo \inst{1}
\and
A. Fuente \inst{3}
\and
H. A. Wootten \inst{4}
\and
E. Chapillon \inst{5,6}
\and
J. Pety \inst{5,2}
\and
D. C. Lis \inst{2}
\and
E. Roueff \inst{2}
\and
B. Commer\c con \inst{7}
\and
A. Ciardi \inst{2}
          }

\institute{
Instituto de F\'isica Fundamental, CSIC, C/ Serrano 123, 28006 Madrid, Spain
\email{nuria.marcelino@csic.es}
\and
Sorbonne Universit\'e, Observatoire de Paris, Universit\'e PSL, \'Ecole normale sup\'erieure, CNRS, LERMA, F-75014, Paris, France
\and
Observatorio Astron\'omico Nacional (OAN,IGN), Apdo 112, E-28803 Alcal\'a de Henares, Spain.
\and
National Radio Astronomy Observatory, 520 Edgemont Road, Charlottesville, VA 22903, USA.
\and
Institut de Radioastronomie Millim\'etrique (IRAM), 300 rue de la Piscine, 38406 Saint Martin d'H\`eres, France.
\and
Laboratoire d'astrophysique de Bordeaux, Univ. Bordeaux, CNRS, B18N, all\'ee Geoffroy Saint-Hilaire, 33615 Pessac, France.
\and
Univ Lyon, Ens de Lyon, Univ Lyon1, CNRS, Centre de Recherche Astrophysique de Lyon UMR5574, F-69007, Lyon, France.
             }

\date{Received ; accepted }



\abstract{The Barnard 1b core shows signatures of being at the earliest stages of low-mass star formation, 
with two extremely young and deeply embedded protostellar objects. Hence, this core is an ideal target to 
study the structure and chemistry of the first objects formed in the collapse of prestellar cores. We present 
ALMA Band 6 spectral line observations at $\sim$0.6$''$ of angular resolution towards Barnard 1b. We have 
extracted the spectra towards both protostars, and used a Local Thermodynamic Equilibrium (LTE) model to 
reproduce the observed line profiles.
B1b-S shows rich and complex spectra, with emission from high energy transitions of complex molecules, 
such as CH$_3$OCOH and CH$_3$CHO, including vibrational level transitions. We have tentatively detected 
for the first time in this source emission from NH$_2$CN, NH$_2$CHO, CH$_3$CH$_2$OH, CH$_2$OHCHO, 
CH$_3$CH$_2$OCOH and both $aGg'$ and $gGg'$ conformers of (CH$_2$OH)$_2$. This is the first detection 
of ethyl formate (CH$_3$CH$_2$OCOH) towards a low-mass star forming region. 
On the other hand, the spectra of the FHSC candidate B1b-N are free of COMs emission. 
In order to fit the observed line profiles in B1b-S, we used a source model with two components: an inner hot 
and compact component (200 K, 0.35$''$) and an outer and colder one (60 K, 0.6$''$). 
The resulting COM abundances in B1b-S range from 10$^{-13}$ for NH$_2$CN and NH$_2$CHO, 
up to 10$^{-9}$ for CH$_3$OCOH.
Our ALMA Band 6 observations reveal the presence of a compact and hot component in B1b-S, with moderate 
abundances of complex organics. These results indicate that a hot corino is being formed in this very young 
Class 0 source.}


\titlerunning{Signatures of an incipient hot corino in B1b-S}

\maketitle
%

\section{Introduction}

In the innermost regions of low-mass Class 0 objects, gas and dust can reach temperatures of $>$100 K, 
at which water and other grain mantle components can evaporate into the gas phase. In particular, the large 
contribution of complex organic molecules (COMs; defined as species with 6 or more atoms) to their 
spectra, makes them similar to hot cores in massive star forming regions, and they are therefore known as 
hot corinos \citep{herbst09,caselli12}.
These regions have small sizes, of $<$100 au, and their study has greatly improved in the last few years 
thanks to interferometers operating at milimeter and submilimeter wavelengths such as NOEMA and ALMA. 
In particular, ALMA observations at unprecedent angular resolution, allow to probe the inner hot regions of 
low-mass protostars and compare with sophisticated MHD models \citep[see, e.g.][]{commercon12,hincelin16}.

Of particular interest are the earliest stages of low-mass star formation: the initial collapse, the formation of 
accretion disks and the launching of protostellar outflows. 
The Barnard 1 cloud is a moderately active star forming region in the Perseus molecular complex, at a 
distance of 230 pc \citep{hirota08}. It contains several dense cores at different stages of star formation, two 
of them, B1a and B1c, hosting class 0 sources associated with high velocity outflows \citep{hatchell05,hatchell07}.
The protostellar core B1b contains three remarkable sources: the infrared source B1b-W, also observed 
recently at millimeter and submillimeter wavelengths \citep{gerin17,cox18}, and two extremely young protostellar 
objects, B1b-N and B1b-S. These two sources have been suggested to be First HydroStatic Core (FHSC) candidates, 
which is the first object to be formed in the collapse of a prestellar core \citep{larson69}. Therefore, this core has 
been a target of multiple studies \citep[see, e.g.][]{pezzuto12,hirano14}. 
Recent interferometric observations with NOEMA and ALMA have shown the incipient star formation 
activity in B1b. \citet{gerin15} observed young and slow outflows driven by both sources, and obtained 
outflow properties consistent with B1b-S being a very early Class 0 and B1b-N a FHSC candidate. ALMA 
continuum emission at 350 GHz at a very high resolution (0.1$''$), indicate the presence of small compact 
structures that were interpreted as nascent disks by \citet{gerin17}. 
Using NOEMA at $\sim4''$, \citet{fuente17} imaged the 
emission of NH$_2$D, which is tracing the pseudo-disk, and detected rotation around B1b-S.
Here we present spectral line ALMA observations at $0.6''$ angular resolution of the two young protostars 
B1b-N and B1b-S.

\section{Observations}

ALMA Cycle 3 observations were performed on 2016 June 21-24, in four 
different tracks of about 1\,hr each. The ALMA 12m array was used in the C40-4 configuration, 
with 36--40 antennas and baselines between 15.1 and 704.1 m. The mean angular resolution achieved 
(interferometric synthesised beam)
at the observed frequencies is $\sim$0.7$\times$0.55$''$ (160$\times$126\,au), and the 
largest scales to which we are sensitive are $\sim$4$''$.
We used the Band 6 receiver centered at 218 and 233 GHz in the lower and upper side band, 
respectively, connected to the ALMA correlator in FDM mode. In total, 13 spectral 
windows were used, one at low resolution ($\sim$40\,km\,s$^{-1}$ and 2\,GHz of bandwidth) 
for the continuum, and 12 windows with spectral resolutions of $\sim$0.16--0.2 
km\,s$^{-1}$ and a 58.5\,MHz bandwidth. 

We performed a 7-point mosaic centered at $\alpha_{J2000}=03^{\rm h} 33^{\rm m} 21.3^{\rm s}$ 
and $\delta_{J2000}=31^\circ 07' 34.0''$, covering an area of 40\,arcsec$^2$ containing B1b-N, 
B1b-S, and the infrared source B1b-W. 
All the fields were observed in scans of $\sim$7\,min duration, 
with 30\,sec scans for phase calibration in-between. Total time on-source is 44\,min per track. 
The phase calibrator used is J0336+3218, located at a distance of $\sim$1.36$^\circ$ of the B1b 
cores. Two other quasars were observed for bandpass (J0237+2848) and flux (J0238+1636) 
calibration. Flux density for J0238+1636 was set to 
$\sim$1.14--1.22\,Jy, depending on the day and frequency, based on close 
flux measurements.
Calibration was performed by the ALMA Pipeline in CASA\footnote{https://casa.nrao.edu/} 4.5.3. 
Images of all spectral windows were also produced in CASA, with a spectral resolution of 0.2 km\,s$^{-1}$ 
and Briggs weighting.
Table~\ref{table:obs} gives the frequency ranges, the final beam and rms of the images.
For analysis purposes we used the GILDAS\footnote{http://www.iram.fr/IRAMFR/GILDAS} software. 
We have extracted the spectra, within one synthesized beam size, towards the positions of 
the B1b-N and B1b-S cores.
Depending on the frequency and the beam for each spectral window, we have converted the 
flux into brightness temperature to analyse the spectra
(K/Jy$\sim$63).

\section{Results}

\begin{figure*}
\includegraphics[width=\hsize]{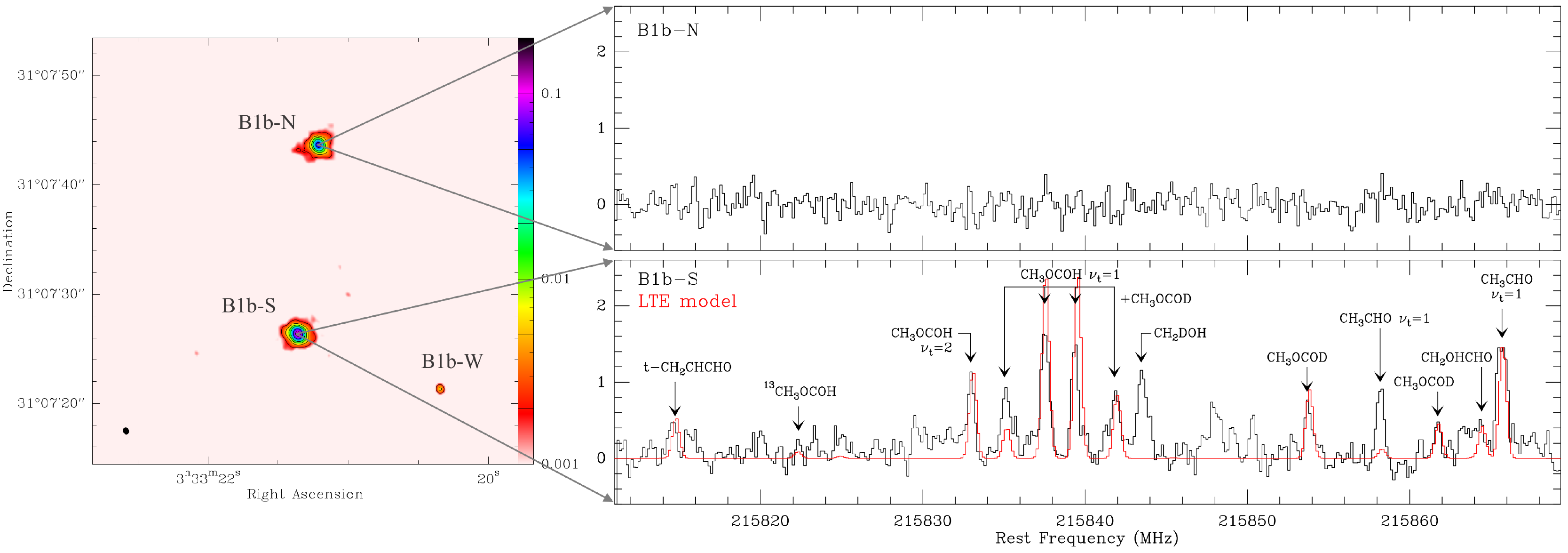}
\caption{
{\it(Left)}: Continuum image at 232~GHz (1.290 mm). The color scale ranges
from 0.001 to 0.2 Jy/beam. The contour levels are indicated
on the color bar. The ALMA beam is plotted on the lower left corner.
{\it(Right)}: Observed spectra between 215810--215870 MHz, towards B1b-N (top) and B1b-S (bottom). 
Brightness temperature scale, in K, is given on the left.
Line features with no labels are unidentified lines. The LTE simulations are overplotted in red for B1b-S.}
\label{fig:figSpectra}
\end{figure*}

The obtained spectra towards the two sources look very different: while B1b-S shows intense and rich spectra, 
B1b-N is mostly free of line emission. 
Right panels in Fig.~\ref{fig:figSpectra} show one of the observed spectral windows, between 215810--215870 MHz, 
towards B1b-N (top) and B1b-S (bottom), and Figs.~\ref{fig:figSpecFull1}-\ref{fig:figSpecFull3} present the full 
spectral coverage. 
In the northern source, we observe lines from $^{12}$CO, $^{13}$CO, C$^{18}$O, $^{13}$CS, DCO$^+$, DCN, 
N$_2$D$^+$, and SO. All of them show extended emission and towards the position of the protostar present complex 
line profiles, which in some cases are only seen in absorption (e.g. $^{12}$CO, and SO). 
One transition of H$_2$C$^{17}$O is marginally detected in emission. 
All these species are also observed in B1b-S, together with 
lines from H$_2$CO, H$_2^{13}$CO, D$_2$CO, CH$_3$OH, $^{13}$CH$_3$OH, CH$_2$DOH, CH$_3$OD 
and several COMs. 
In total, we have detected 190 lines towards B1b-S. Of those, 113 spectral features have been assigned to 155 
transitions from 36 species (including isotopologues and vibrational states).
Table~\ref{table:fits} lists all assigned transitions and their associated species in B1b-S, with line parameters obtained 
from Gaussian line fits. Line identification was carried out using 
MADEX\footnote{\texttt{https://nanocosmos.iff.csic.es/$?$page\_id=1619}} \citep{madex}.
The other 77 lines for which we could not find reasonable carriers are labeled as unidentified ($\sim$40\% 
of the total number of lines).  Of those, 16 have also been observed in Orion KL, that is only 21\% of 
the unidentified lines. All transitions listed in Table~\ref{table:fits} are above the 3$\sigma$ level 
(with $\sigma=rms \sqrt{Dvdv}$, $Dv$ the linewidth, and $dv$ the spectral resolution)
, while U lines were considered above 4$\sigma$.
Several lines, mainly from the most abundant species and low energy transitions, show extended emission when 
looking at the maps and present complex line profiles towards the position of B1b-S (lines in absorption, 
inverse P-Cygni profiles, line wings, etc.). These lines are indicated in Table~\ref{table:fits} by CP or 
Ext (for complex profile and 
extended, respectively), and we are not including them in the analysis below. 
While we have identified all the lines in the observed spectra (see Figs.~\ref{fig:figSpecFull1}-\ref{fig:figSpecFull3} 
and Table~\ref{table:fits}), the analysis of the extended emission is outside the scope of this paper. Furthermore 
it is likely that a significant fraction of the flux is resolved out in the ALMA 12m-array observations.

In the following we focus on the emission from COMs in the protostar B1b-S.
We have detected multiple transitions of acetaldehyde (CH$_3$CHO) and methyl formate (CH$_3$OCOH), including emission 
from their deuterated substitutions and vibrational states. 
We have also observed emission from isocyanic acid (HNCO), cyanamide (NH$_2$CN), formamide (NH$_2$CHO), vinyl cyanide 
(CH$_2$CHCN), propenal (t-CH$_2$CHCHO), ethanol (t-CH$_3$CH$_2$OH), dimethyl ether (CH$_3$OCH$_3$), ethyl formate 
(t-CH$_3$CH$_2$OCOH), glycolaldehyde (CH$_2$OHCHO) and both $aGg'$ and $gGg'$ conformers of ethylene glycol 
((CH$_2$OH)$_2$). However, some of these should be considered tentative detections. 
In particular, we have detected only one unblended line of HNCO, NH$_2$CN, NH$_2$CHO, t-CH$_2$CHCHO, CH$_2$OHCHO, 
t-CH$_3$CH$_2$OH, and CH$_3$OCH$_3$. 
Of the 16 transitions observed from the ethylene glycol conformers, 8 are blended with transitions from 
other species. We have clearly detected 3
unblended lines from CH$_3$CH$_2$OCOH, which has only been previously detected towards Sgr B2, Orion KL, 
and the hot core in W51 \citep{belloche09,tercero13,rivilla17}. 
We would like to point out the large number of unassigned lines and the high level of blending and complexity in 
the spectra (see Figs.~\ref{fig:figSpecFull1}-\ref{fig:figSpecFull3} and Table~\ref{table:fits}), like that observed typically in hot cores 
and hot corinos. The emission of all these complex species is unresolved in the ALMA images. We have checked that 
at one beam separation from the position of the B1b-S protostar the spectra are free of COMs emission.

\section{Column densities and abundances}

We have used MADEX \citep{madex} to reproduce the observed spectra of COMs in B1b-S, and to obtain 
column densities and kinetic temperatures of the emitting region. We assume local thermodynamic equilibrium 
(LTE) conditions, since the densities at the center of the core are expected to be very high \citep{gerin17}.
We have focused on CH$_3$OCOH and CH$_3$CHO, and their isotopic substitutions and vibrationally excited 
states, for which many transitions are available, to obtain the source model which was then used for the 
other observed species. 

Methyl formate is the species for which the largest number of transitions were identified. In total 24 
transitions from both A and E species were observed (18 unblended), with energy levels between 100 and 
500 K, allowing to constrain the properties of the emitting region from the LTE model. 
We run a set of models using one source component of 0.6$''$ in size (the angular resolution of the ALMA band 6 
observations), and we modified the temperature and column density to fit the observed spectra. However, 
it was difficult to fit all the lines with only one component.
The problem became more evident when using the deuterated substitutions, CH$_3$OCOD and CH$_2$DOCOH, and 
the vibrationally excited states $\nu_t=1$ and $\nu_t=2$.
We have detected 12 transitions from CH$_3$OCOD and 4 transitions from CH$_2$DOCOH (3 and 2 respectively, 
are blended with other species), with upper energy levels between 90 and 170 K. 
In both cases the model 
gave better results when decreasing the temperature with respect to the best CH$_3$OCOH model. 
On the other hand, the vibrationally excited lines (13 unblended transitions from $\nu_t=1$, and 1 transition 
from $\nu_t=2$), supported the high temperature model. 
Therefore 
we built a 
model of the source with two components. One warm and compact component at 200 K and a size of 0.35$''$ 
\citep[that of the continuum compact emission obtained by ALMA Band 7 observations,][]{gerin17}, 
and another larger (0.6$''$, approximately the angular resolution of the Band 6 observations) and colder 
component at 60 K. 
Figure~\ref{fig:figComparaModel} shows two examples of the best one component models obtained using the main 
CH$_3$OCOH (200 K), and the deuterated methyl formate lines (60 K), together with the final two component model. 
While some species give reasonable agreement with observations using either model (mostly those which 
observed transitions have similar energy levels), there are others (besides the ones discussed above) 
that fail in one of the models (e.g. HNCO and t-CH$_3$CH$_2$OCOH). 
We have checked that using other temperatures (between 50--250 K) in the one component models does not 
reproduce satisfactorily the observations. 
For the vibrationally excited states,
we have only 
used the compact component at 200 K, since the low temperature one does not contribute much to the line 
intensities. Indeed, energy levels of the observed transitions range between 300 and 500 K, supporting 
this scenario. 
We have also observed 5 transitions of $^{13}$CH$_3$OCOH, with energies $\leq$200 K and intensities of 
the order of 500 mK or below. Since the lines are weak, any of the models give reasonable agreement with 
the observations.
Acetaldehyde is the other species for which many lines are observed. We have detected 15 transitions 
(10 unblended) from the main species, with energies between 100--300 K. 
We have also detected one transition arising from the deuterated substitution CH$_3$CDO, and 4 transitions 
(2 of them blended) from the vibrational state $\nu_t=1$, with energy levels of $\sim$300--400 K. 
We used the same model as methyl formate, except for CH$_3$CHO $\nu_t=1$, for which only the compact and 
hot component was considered.
The model allows to reproduce all transitions reasonably well, except for a few lines that are overestimated: 
two CH$_3$OCOH transitions (Fig.~\ref{fig:figSpecFull1}, second panel), three CH$_3$OCOH $\nu_t=1$ lines 
(top panel in Fig.~\ref{fig:figSpecFull1}, and second panel in Fig.~\ref{fig:figSpecFull3}), and 2 lines 
from CH$_3$CHO (see Fig.~\ref{fig:figSpecFull3}, second and third panels). 
These are transitions with the lowest energies ($\sim$100 K) and the highest line strengths and Einstein 
coefficients, and may consequently have high opacities. 
We estimate the error in the fit to be $\sim$50\%.

This simple two component model simulates the temperature gradient that should exist in the source. It is 
possible that a more sophisticated model will reproduce the lines better, however the limited 
angular resolution and signal-to-noise ratio of the current observations do not allow a well-constrained model. 
In particular, the angular resolution 
is not sufficient to separate the emission from the compact and hot component at 200\,K 
from that of the lukewarm gas at 60\,K.
For example, it is possible that the size of 
the hot component could be smaller, and therefore we underestimate the column densities.

For the other COMs, we have used the same two component model as for methyl formate and acetaldehyde 
for consistency. 
There are two exceptions: the observed lines from $^{13}$CH$_3$OH and HNCO arise from very 
excited energy levels ($\sim$ 600 K and 700 K respectively). Hence we have assumed that only the hot and 
compact component contributes to the detected emission.
As mentioned above, we have excluded from the analysis all the species with lines showing complex 
profiles and extended emission. 
We have attempted to run the LTE models for CH$_3$OH, H$_2$CO and their isotopologues, which do not show absorption 
or very complex profiles. All the observed lines of H$_2$CO, H$_2^{13}$CO, H$_2$C$^{17}$O and D$_2$CO (only 
one transition each) have low $E_{up}$ values and show extended emission. Therefore these lines will suffer from 
flux loss due to filtering from the interferometer, and the obtained column densities are strongly underestimated. 
For CH$_3$OH and CH$_2$DOH we observed more than one transition (2 and 6 respectively), 
but they span a considerable $E_{up}$ range, which makes a simultaneous fit difficult. 
In particular, lines with low energies and extended emission are significantly overestimated. We have 
observed only one transition from CH$_3$OD (also extended) and $^{13}$CH$_3$OH. 
We have only kept the LTE result for the latter which does not present any extended emission.
Lines from DCO$^+$, DCN, C$^{18}$O, SO, $^{13}$CO, CO, $^{13}$CS, and N$_2$D$^+$, are seen in absorption or 
show very complex profiles, and we have not attempted to fit them with the LTE model. 
For all these species, 
we show the line identification in Fig.~\ref{fig:figSpectra} and 
Figs.~\ref{fig:figSpecFull1}-\ref{fig:figSpecFull3}.

\begin{table}
\caption{Results from LTE model.}
\label{table:results}
\centering
\begin{tabular}{lcc}
\hline\hline
 Species         &  $N$ (cm$^{-2}$)       & $X/H_2$\tablefootmark{a} \\
\hline
NH$_2$CN   & 1.0 ; 1.0 10$^{12}$ & 7.1 ; 9.1 10$^{-14}$ \\
HNCO &  3.0 ; --- 10$^{14}$   &  2.1 ; --- 10$^{-11}$ \\
$^{13}$CH$_3$OH &  5.0 ; --- 10$^{15}$   &  3.6 ; --- 10$^{-10}$ \\
CH$_3$CHO  &  8.0 ; 1.6 10$^{14}$   & 5.7 ; 1.4 10$^{-11}$ \\
CH$_3$CHO $\nu_t=1$ & 4.0 ; --- 10$^{14}$ & 2.8 ; --- 10$^{-11}$ \\
CH$_3$CDO  & 4.0 ; 4.0 10$^{13}$   &  2.8 ; 3.6 10$^{-12}$ \\
NH$_2$CHO  &  4.0 ; 4.0 10$^{12}$   &  2.8 ; 3.6 10$^{-13}$ \\
CH$_3$OCH$_3$   &  1.0 ; 1.0 10$^{16}$   &   7.1 ; 9.1 10$^{-10}$ \\
$t$-CH$_3$CH$_2$OH  &  2.0 ; 2.0 10$^{14}$   &  1.4 ; 1.8 10$^{-11}$ \\
CH$_2$CHCN   &  4.0 ; 2.0 10$^{13}$   &  2.8 ; 1.8 10$^{-12}$ \\
$t$-CH$_2$CHCHO   &  3.0 ; 2.0 10$^{13}$   &  2.1 ; 1.8 10$^{-12}$ \\
CH$_2$OHCHO   &  5.0 ; 3.0 10$^{14}$   &  3.6 ; 2.7 10$^{-11}$ \\
CH$_3$OCOH  & 1.5 ; 0.5 10$^{16}$ & 1.1 ; 0.4 10$^{-9}$ \\
CH$_3$OCOH $\nu_t=1$   &  2.0 ; --- 10$^{15}$ & 1.4 ; --- 10$^{-10}$ \\
CH$_3$OCOH $\nu_t=2$   &  8.0 ; --- 10$^{14}$  &  5.7 ; --- 10$^{-11}$ \\
$^{13}$CH$_3$OCOH  &  3.0 ; 3.0 10$^{14}$  &  2.1 ; 2.7 10$^{-11}$ \\
CH$_3$OCOD  & 3.0 ; 5.0 10$^{14}$   &   2.1 ; 4.5 10$^{-11}$ \\
CH$_2$DOCOH  &  0.4 ; 1.0 10$^{15}$   &  2.8 ; 9.1 10$^{-11}$ \\
aGg'-(CH$_2$OH)$_2$  &  2.0 ; 0.8 10$^{14}$ & 1.4 ; 0.7 10$^{-11}$ \\
gGg'-(CH$_2$OH)$_2$   & 9.0 ; 6.0 10$^{13}$ & 6.4 ; 5.4 10$^{-12}$ \\
$t$-CH$_3$CH$_2$OCOH  &  4.6 ; 0.8 10$^{14}$   &  3.3 ; 0.7 10$^{-11}$ \\
\hline
\end{tabular}
\tablefoot{
First values correspond to the compact and hot component (200 K, 0.35$''$), 
second values to the extended and cold one (60 K, 0.60$''$). 
For vibrationally excited transitions, HNCO and $^{13}$CH$_3$OH, 
only the hot component was considered.
\tablefoottext{a}{Using N(H$_2$)=1.4 10$^{25}$ cm$^{-2}$ at 0.35'', 
and N(H$_2$)=1.1 10$^{25}$ cm$^{-2}$ at 0.60'' (see text).}
}
\end{table}

Table~\ref{table:results} shows, for each source component, the obtained column densities and molecular abundances with respect to H$_2$. 
In order to compute the abundances we have estimated the molecular hydrogen column densities for both components. 
Based on previous observations of dust continuum emission with ALMA at 850 $\mu$m and JVLA at 8-10 cm 
(those with the best angular resolution, see appendix~\ref{app:continuum}) and assuming dust temperatures from 12 to 50 K,
we have computed mean column densities of 
N(H$_2$)=1.4 10$^{25}$ cm$^{-2}$ and 1.1 10$^{25}$ cm$^{-2}$ for source sizes of 0.35$''$ and 0.6$''$, respectively. 
While the dust emission is optically thick at 350 GHz on size scales smaller than 0.3$''$, the 8\,cm emission is 
optically thin. Then it is possible the H$_2$ column density estimates could be affected by the opacity, in 
particular at 0.35'' more than at 0.6''. Another important source of error comes from the unknown dust properties, 
since the standard opacity law may be different if dust grains have grown. All these factors could result in 
uncertainties in the column density estimates up to a factor 2.

\section{Discussion}

Previous spectral studies with the IRAM 30m telescope already pointed out the presence of complex organics in B1b. 
\citet{cerni12} reported the detection of several COMs, including CH$_3$CHO, CH$_3$OCOH, and CH$_3$OCH$_3$, 
also observed with ALMA. These detections are based on low energy transitions, $<$25 K, and because of the 
large 30m beam at 3mm ($30-21''$) the emission should be arising from the cold envelope. 
Table~\ref{table:compare} shows the abundances derived from the ALMA data, together with those obtained 
in the envelope from the IRAM 30m survey at 3mm. 
Except for CH$_2$CHCN and HNCO, there is a general trend of COMs being enhanced in the inner core of B1b-S with respect to 
the colder protostellar envelope, with methyl formate showing the highest increase. 
This is possibly related to the evaporation of grain mantles due to the increased dust temperature caused by the heating from the embedded protostar.
A similar increase in abundance is seen in other hot corinos with respect to the values observed in their envelopes. In IRAS 16293-2422, \citet{jaber14} obtain higher abundances in the inner warm envelope with respect to the outer cold envelope, and these values are lower than those observed at smaller scales using interferometers, i.e. closer to the protostar \citep{jorgensen12,coutens17}.
The range of temperatures obtained for B1b-S cores in the LTE model, from 60\,K up to at least 200\,K in a size of 0.35'', and the narrow 
linewidths are consistent with thermal evaporation. Other mechanisms could also be at work, such as accretion shocks in the disk or launching of the outflow. However we can not conclude about the origin of COMs since we lack angular resolution.
Indeed, it is possible that the size of the compact component in the model (0.35'') is smaller, 
since we do not resolve the emission from the highest energy transitions. 

In the ALMA data we have also detected the singly deuterated isotopologues of acetaldehyde and methyl formate, whose 
abundances decrease at the inner hot component with respect to the component at 60 K, with D/H 
a factor 5 higher in the latter (see Table~\ref{table:compare}).  
These deuterated species are not detected with the IRAM 30m, possibly because the main species are weak and less abundant 
in the envelope. However high deuteration levels are observed for other species, e.g. strong HDCO lines were detected by 
\citet{marce05} with a D/H ratio of 0.145. 
This value is similar to those derived in the external B1b-S core for CH$_3$OCOD and CH$_2$DOCOH (D/H of 0.1 and 0.2, respectively), 
and smaller than CH$_3$CDO/CH$_3$CHO = 0.25. 
High deuterium fractionation is also observed in hot corinos, while hot cores show lower deuterium abundances 
\citep[see][and references within]{caselli12}.

The comparison with other hot corinos is not easy since we have a limited number of lines, in contrast 
to well-known sources that have been more extensively observed with ALMA and NOEMA, such as IRAS 16293-2422 
\citep{jorgensen16}, NGC1333 IRAS 2A and 4A \citep{taquet15,sepulcre17}. In general, we find a similar 
inventory of COMs but with lower abundances in B1b-S. We are going to discuss only some trends and 
ratios between them. For example, one
similarity with hot corinos are the higher abundances of O-bearing COMs with respect to N-bearing COMs. 
We observe a similar trend in B1b-S, where the highest abundances are obtained for CH$_3$OCOH, CH$_3$OCH$_3$, 
and CH$_3$CHO, while NH$_2$CN and NH$_2$CHO show the lowest abundances. 
Cyanamide has been recently detected towards the low-mass protostars IRAS 16293-2422 and NGC1333 IRAS2A 
\citep{coutens17}, with 
observed ratios NH$_2$CN/NH$_2$CHO of 0.2 and 0.02, respectively. These values are in the range of those observed 
towards the molecular clouds in the Galactic Center but lower than in Orion KL \citep{coutens17}. 
We obtain in B1b-S an abundance ratio of 0.25, similar to IRAS 16293-2422. 
Of the three possible isomers of C$_2$H$_4$O$_2$, we have detected CH$_3$OCOH and CH$_2$OHCHO. 
The observed CH$_3$OCOH/CH$_2$OHCHO ratio in B1b-S, $\sim$20, is similar to that observed in the low-mass 
protostars in NGC 1333 and IRAS 16293-2422 \citep{taquet15,jorgensen12}. 
Acetic acid (CH$_3$COOH) is the most stable but the least abundant of the three isomers \citep{lattelais10}.
It has been observed in IRAS 16293-2422 with
a ratio with respect to glycolaldehyde of $\sim$11 \citep{jorgensen16}, consistent with the 5--15 upper limits in B1b-S. 
Glycolaldehyde and its corresponding alcohol, ethylene glycol, show similar abundances in B1b-S, 
slightly higher for CH$_2$OHCHO. This is 
in contrast to
other hot cores and hot corinos 
\citep{fuente14,jorgensen16,favre17}. However, since detections of these species are based on few lines, 
in particular for glycolaldehyde, 
column densities may be not well constrained and it is difficult to make definite conclusions.
We have detected CH$_3$CH$_2$OCOH for the first time in a low-mass protostar, with a relatively high abundance of 10$^{-11}$.
This detection shows that the molecular complexity is high in young hot corinos and other species with a 
larger number of atoms could be present. It is noteworthy that this species has not 
been previously observed in the well-known hot corinos IRAS 16293-2422 and NGC133IRAS4A. 
We have checked in the ALMA spectral line survey of IRAS 16293-2422 \citep[PILS,][]{jorgensen16}, and no clear 
features are seen, although the observed frequencies are different. 
It is possible that the high level of line confusion and wide linewidths prevent the detection of 
weak lines in these more evolved sources. 

Observed line widths in B1b-S are typically narrow, $\sim$1 km\,s$^{-1}$ (see Table~\ref{table:fits}), in contrast 
to the wide line profiles in hot cores and hot corinos ($\sim$3-5 km\,s$^{-1}$).  
Besides all lines from COMs are found in emission, with no P-Cygni or absorption profiles.
Therefore there is no signature of rotation or infall in the disk/core from COMs. 
The orientation of the disk should not prevent the detection of rotation, following the observations of the outflows 
in B1b-S \citep{gerin15}. Actually, using NOEMA data, \citet{fuente17} detect rotation using NH$_2$D. 
The absence of detected rotation could be due to the too low angular
resolution since the COM emission is unresolved. In addition, although
the sensitivity is excellent, these observations are not deep enough to
detect weak features at the 0.3--0.6 K level (see rms in Table~\ref{table:obs}).
Hence weak line wings tracing rotation may have been missed.

\section{Conclusions}

We present new ALMA Band 6 data towards the protostellar system B1b in Perseus. 
We have detected 
a rich and complex spectra in B1b-S, with 190 spectral features. Of those, 99 lines are assigned 
to transitions from complex molecules, while 77 remain unidentified. 
Previous NOEMA observations do not show any emission from COMs, with spatial resolutions between 
2$''$ and 5$''$ \citep{gerin15,fuente17}. These data suggest that embedded in the B1b-S core a 
very young and compact object is already warming up its most immediate surroundings, where 
molecules from grain mantles are evaporating.
On the other hand, B1b-N is almost free of line emission. The non detection of complex species 
supports the scenario in which B1b-N is at an earlier stage of evolution than B1b-S. Another 
possibility is that the more active outflow in B1b-S, with higher velocities and more complex 
profiles (with evidences of bullets), could contribute to a more efficient release of COM in 
the gas phase in B1b-S, as compared to B1b-N.
Luminosity outbursts due to episodic accretion in B1b-S could also increase the dust temperature 
and produce the evaporation of COMs from grain mantles or trigger the formation of daughter 
molecules in the gas phase \citep{taquet16}. The origin of COMs in B1b-S remains unclear since 
the emission is not resolved.

While B1b-S presents a rich spectra similar to other hot corinos, observed 
line widths are typically narrower and do not show any signature of rotation or infall. 
This could be due to the limited angular resolution and signal-to-noise ratio. 
These observations confirm that hot corinos are a common feature
in class 0 protostars, but require sub arcsec angular resolution and high
sensitivity reaching a few mJy/beam to be detected. Higher angular
resolution observations of B1b-S with better sensitivity are desirable 
to resolve the emission from COMs, and better constrain 
molecular abundances, the size and the structure of the emitting region.

\begin{acknowledgements}
We thank the referee for a careful reading of the manuscript and useful comments 
which have improved the paper significantly.
This paper makes use of the following ALMA data: ADS/JAO.ALMA\#2015.1.00025.S. 
ALMA is a partnership of ESO (representing its member states), NSF (USA) and NINS (Japan), 
together with NRC (Canada), MOST and ASIAA (Taiwan), and KASI (Republic of Korea), in 
cooperation with the Republic of Chile. The Joint ALMA Observatory is operated by ESO, AUI/NRAO and NAOJ.
We acknowledge funding support from the European Research Council (ERC Grant 610256: NANOCOSMOS), 
the ANR project IMOLABS, and the Spanish MINECO under projects AYA2016-75066-C2-2-P and AYA2012-32032.
This work was partly supported by the Programme National "Physique et Chimie du Milieu Interstellaire"
(PCMI) of CNRS/INSU with INC/INP, co-funded by CEA and CNES.
\end{acknowledgements}

%
%

\begin{appendix} 

\section{Continuum results}
\label{app:continuum}

The frequency setup included a broad spectral window, in order to detect
the continuum
emission of the B1b sources, and to facilitate the bandpass and phase
calibration on
the reference quasars.
We present in the left panel in Fig.~\ref{fig:figSpectra} the resulting continuum image. The
three Barnard 1b protostars are detected, B1b-N, B1b-S and even the weak
source
B1b-W. Table \ref{tab:cont}
provides a  compilation of the spectral energy distribution for these
three sources.
Given the extended nature of the sources the listed values depend
on the angular resolution of the observations. It is therefore important
to compare data measured at the same angular resolution.

\begin{table*}
\caption{Observed fluxes at different wavelengths for the B1b sources.}
\label{tab:cont}
\centering
\begin{tabular}{lcccccl}
\hline\hline
$\lambda$  & B1b-N           & B1b-S           & B1b-W           & Noise          & Beam  & Refs. \\
($\mu$m)   & (Jy)            & (Jy)            & (Jy)            & (Jy)           &       & \\
\hline
24         & $<$ 2 10$^{-4}$ & $<$ 2 10$^{-4}$ & 0.17            & 0.02           & 12    & 1, 2 \\
70         & $<$ 0.05        & 0.22            & 0.36            & 0.15           & 7     & 1, 2, 3 \\
100        &  0.6            & 2.29            & 1.0             & 0.3            & 8.5   & 2 \\
160        &  3.2            & 9.1             & $<3$            & 1              & 8.9   & 2 \\
250        &   9.5           & 14.4            & $<5$            & 1              & 14.3  & 2 \\
350        &  12.7           & 16.9            & $<10$           & 3              & 18.4  & 2 \\
350        &   6.0           & 7.2             & $<10$           & 0.3            & 20    & 3 \\
450        &  13.5           & 19.1            & $<6$            & 4              & 11    & 2 \\
500        &  14.6           & 15.8            & $<6$            & 7              & 34.3  & 2 \\
850        &  3.3            & 3.8             & $<3$            & 1.3            & 15    & 2 \\
850        &  1.03           & 1.24            & $<1$            & 0.03           & 15    & 3 \\
850        & 0.24            & 0.48            & $<1$            & 0.01           & 0.3   & 4 \\
850        &  0.73           & 0.89            & $<1$            & 0.01           & 1     & 4 \\
870        &  --             &  --             & 0.0194          & 1.4 10$^{-4}$  & 0.3   & 5 \\
1057       & 0.31            & 0.47            & $<0.3$          & 0.015          & 3     & 3 \\
1290       &  0.12           &  0.23           & 6.4 10$^{-3}$   & 1 10$^{-3}$    & 0.6   & 6 \\
1300       & 0.19            & 0.34            & $<0.2$          & 0.01           & 5     & 3 \\
1300       & 0.249           & 0.375           & --              & 0.011          & 4     & 7 \\
2070       & 0.077           & 0.096           & $<0.04$         & 4 10$^{-3}$    & 2.2   & 8 \\
3300       & 0.026           & 0.031           & $<0.02$         & 3.5 10$^{-3}$  & 3     & 3 \\
3660       & 0.018           & 0.016           & $<$ 1 10$^{-3}$ & 1 10$^{-3}$    & 3.9   & 8 \\
7000       & 1.9 10$^{-3}$   & 1.7 10$^{-3}$   & $<$ 2 10$^{-3}$ & 2.8 10$^{-4}$  & 0.5   & 3 \\
8000       & 1.25 10$^{-3}$  & 0.91 10$^{-3}$  & 1 10$^{-4}$     & 4 10$^{-5}$    & 0.2   & 9 \\
10000      & 6.8 10$^{-4}$   & 5.5 10$^{-4}$   & 8 10$^{-5}$     & 3 10$^{-5}$    & 0.2   & 9 \\
\hline
\end{tabular}
\tablebib{
(1)~\citet{evans09};
(2) \citet{pezzuto12}; 
(3) \citet{hirano14}; 
(4) \citet{gerin17};
(5) \citet{cox18};
(6) This work; 
(7) \citet{pokhrel18};
(8) \citet{gerin15}; 
(9) \citet{tobin16}.
}
\end{table*}

\section{Additional figures and tables}

\begin{table}
\caption{Summary of the observed spectral line windows.}
\label{table:obs}
\centering
\begin{tabular}{c c c c c}
\hline\hline
 Frequency range   &     beam & PA\tablefootmark{a} & $\sigma$F\tablefootmark{b} & rms\tablefootmark{c} \\
          (MHz)           &     ($''$)   &          ($^\circ$)         &          (mJy/beam)                 & (mK) \\
\hline
 215811 -- 215869  &  0.74$\times$0.57 & 23 & 3.29 & 128 \\
 216083 -- 216142  &  0.74$\times$0.57 & 23 & 3.64 & 136 \\
 217076 -- 217134  &  0.73$\times$0.57 & 22 & 3.52 & 184 \\
 217209 -- 217268  &  0.73$\times$0.58 & 16 & 3.78 & 153 \\
 218432 -- 218490  &  0.74$\times$0.56 & 26 & 3.71 & 216 \\
 219531 -- 219590  &  0.70$\times$0.57 & 22 & 3.86 & 196 \\
 219899 -- 219957  &  0.69$\times$0.57 & 19 & 4.14 & 169 \\
 220367 -- 220426  &  0.69$\times$0.57 & 18 & 4.76 & 250 \\
 230519 -- 230577  &  0.67$\times$0.54 & 17 & 4.46 & 193 \\
 231192 -- 231250  &  0.67$\times$0.53 & 15 & 4.57 & 194 \\
 231293 -- 231351  &  0.59$\times$0.44 & 16 & 6.17 & 414 \\
 231381 -- 231439  &  0.68$\times$0.53 & 16 & 4.04 & 184 \\
\hline
\end{tabular}
\tablefoot{
\tablefoottext{a}{Beam Position angle.}
\tablefoottext{b}{rms noise level in the images.}
\tablefoottext{c}{rms noise level measured in the B1b-S spectra.}}
\end{table}

\begin{figure*}
\includegraphics[width=\hsize]{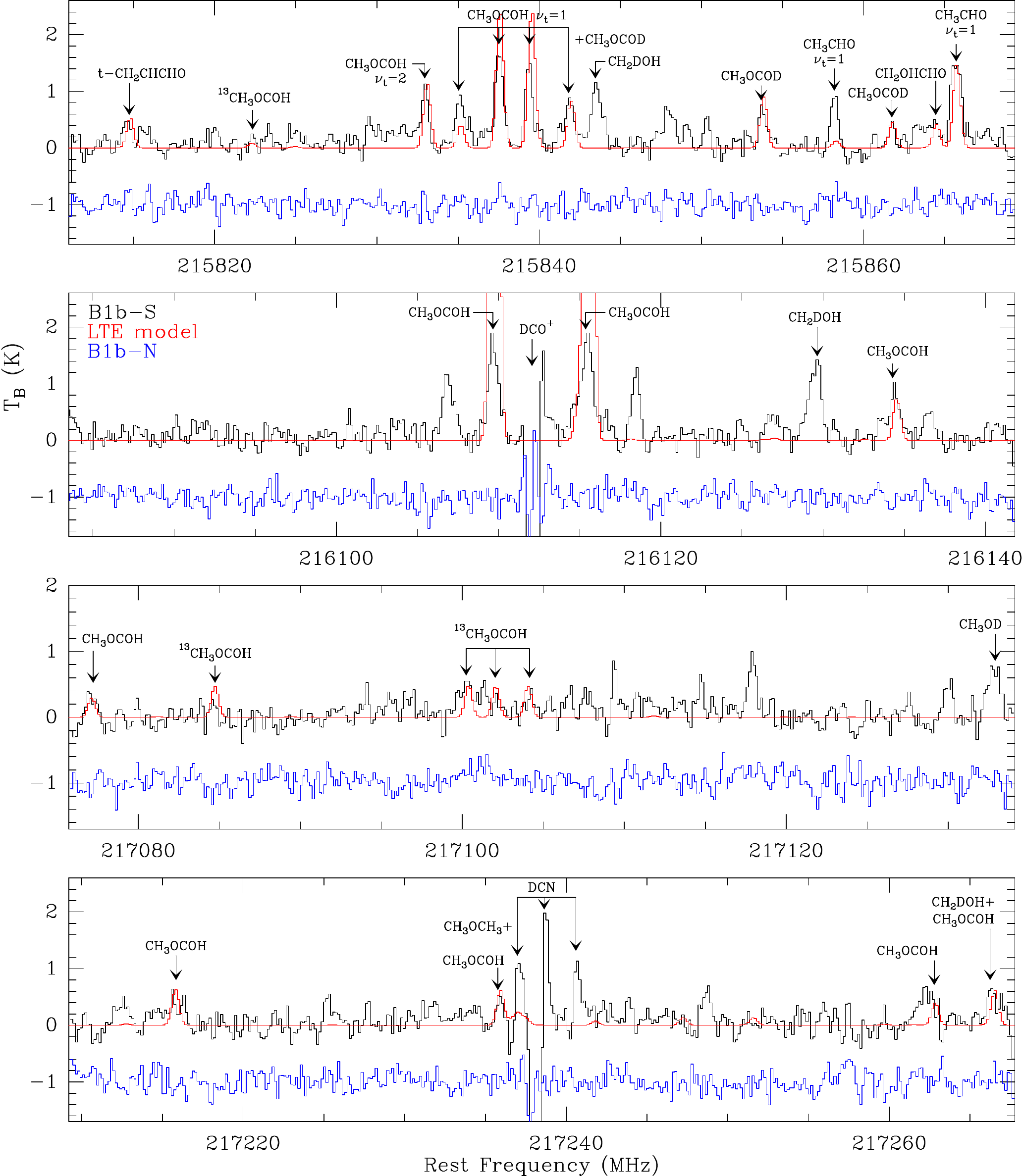}
\caption{Observed spectra towards B1b-S (black) and B1b-N (blue), 
between 215810--217268 MHz. 
Line features with no labels are unidentified lines. 
The LTE simulations for B1b-S are overplotted in red.}
\label{fig:figSpecFull1}
\end{figure*}

\begin{figure*}
\includegraphics[width=\hsize]{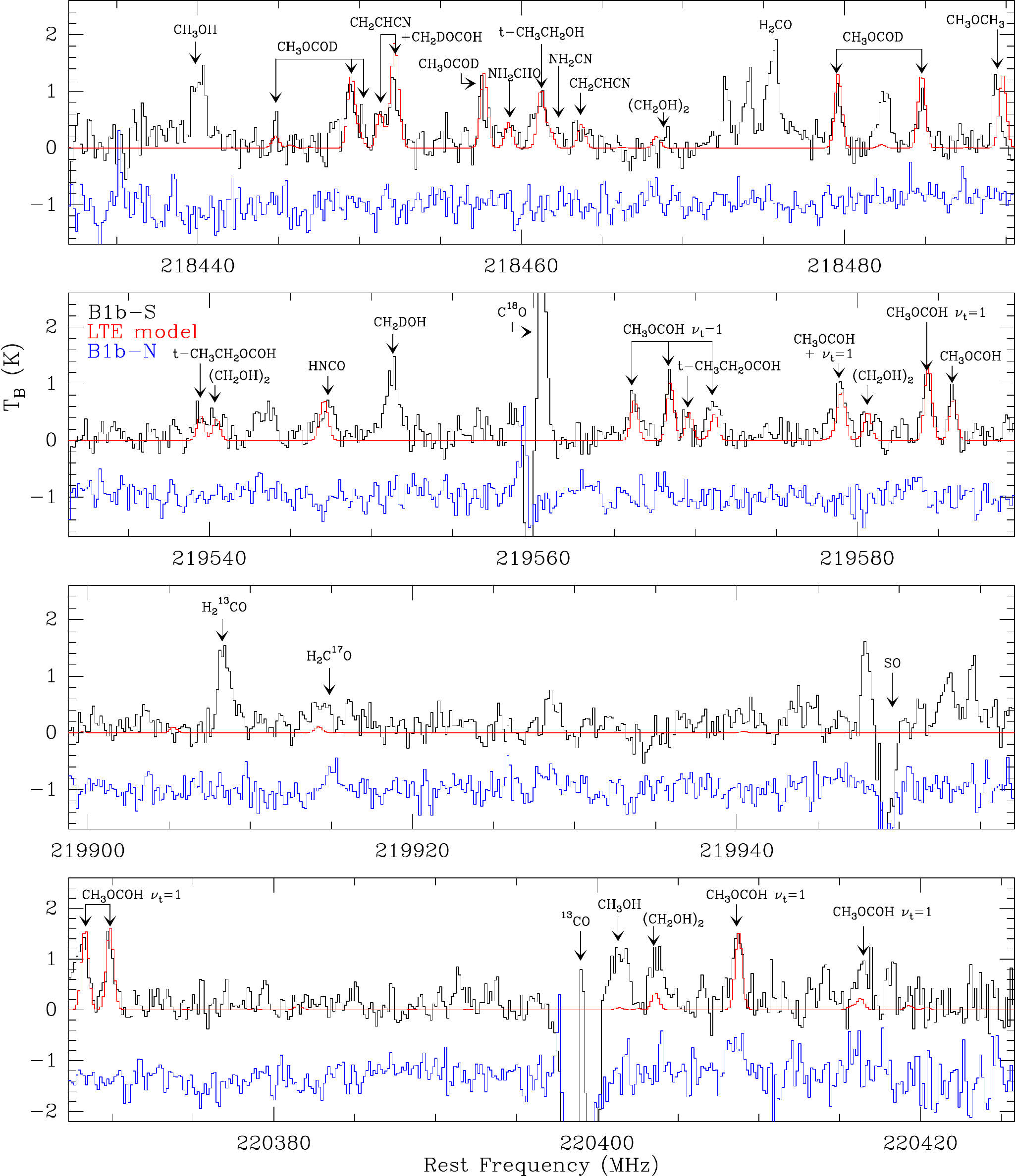}
\caption{Observed spectra towards B1b-S (black) and B1b-N (blue), 
between 218432--220426 MHz. 
Line features with no labels are unidentified lines. 
The LTE simulations for B1b-S are overplotted in red.}
\label{fig:figSpecFull2}
\end{figure*}

\begin{figure*}
\includegraphics[width=\hsize]{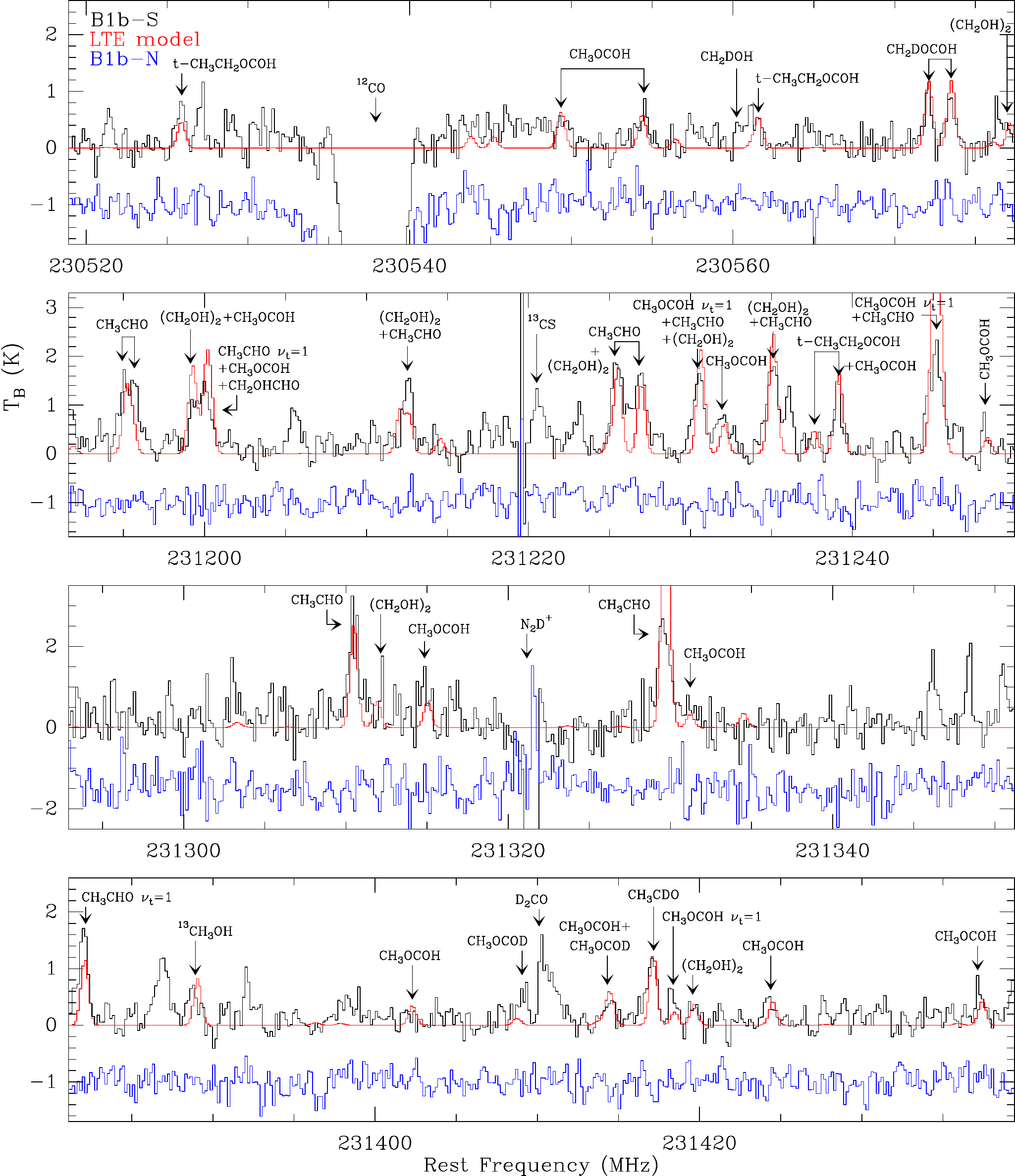}
\caption{Observed spectra towards B1b-S (black) and B1b-N (blue), 
between 230519--231440 MHz. 
Line features with no labels are unidentified lines. 
The LTE simulations for B1b-S are overplotted in red.}
\label{fig:figSpecFull3}
\end{figure*}

\begin{figure*}
\includegraphics[width=\hsize]{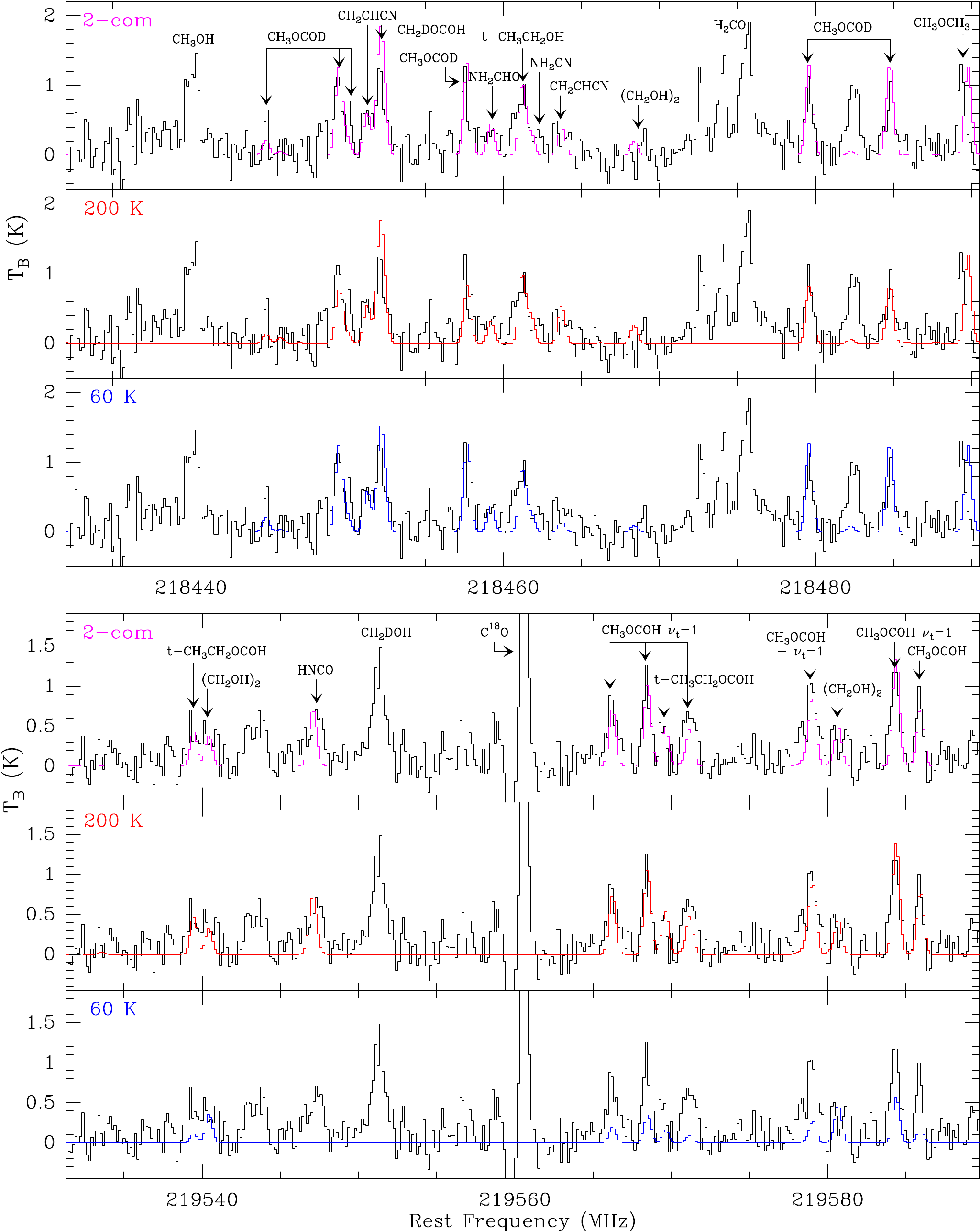}
\caption{Best LTE models using only one source component for two spectral windows. 
Middle panels for each frequency range show the results for the 200 K model in red, 
while bottom panels show in blue the 60 K model. Top panels in purple show the final
2 component model.}
\label{fig:figComparaModel}
\end{figure*}

\longtab[2]{
\begin{landscape}
\begin{longtable}{cllrrcllllc}
\caption{\label{table:fits}
Observed transitions in B1b-S. Line parameters are obtained from Gaussian fits.
Rest frequencies and spectroscopic data are taken from the MADEX catalogue \citep{madex}.}\\
\hline\hline
Obs. Freq. & \multicolumn{1}{c}{Molecule \& Transition}  & \multicolumn{1}{c}{Rest Freq.} & \multicolumn{1}{c}{$E{up}$} & \multicolumn{1}{c}{$S_{ij}$} & $A_{ij}$   & \multicolumn{1}{c}{$\int T_{\rm B} dv$} & \multicolumn{1}{c}{$V_{\rm LSR}$} & \multicolumn{1}{c}{$\Delta$v} & \multicolumn{1}{c}{$T_{\rm B}$} & Notes \\
(MHz)     &         & \multicolumn{1}{c}{(MHz)} & \multicolumn{1}{c}{(K)}  &   & \multicolumn{1}{c}{(s$^{-1}$)} &  \multicolumn{1}{c}{(K km s$^{-1}$)}  & \multicolumn{1}{c}{(km s$^{-1}$)} & \multicolumn{1}{c}{(km s$^{-1}$)} & \multicolumn{1}{c}{(K)}      &      \\
\hline
\endfirsthead
\caption{continued.}\\
\hline\hline
Obs. Freq. & \multicolumn{1}{c}{Molecule \& Transition}  & \multicolumn{1}{c}{Rest Freq.} & \multicolumn{1}{c}{$E{up}$} & \multicolumn{1}{c}{$S_{ij}$} & $A_{ij}$   & \multicolumn{1}{c}{$\int T_{\rm B} dv$} & \multicolumn{1}{c}{$V_{\rm LSR}$} & \multicolumn{1}{c}{$\Delta$v} & \multicolumn{1}{c}{$T_{\rm B}$} & Notes \\
(MHz)     &         & \multicolumn{1}{c}{(MHz)} & \multicolumn{1}{c}{(K)}  &   & \multicolumn{1}{c}{(s$^{-1}$)} &  \multicolumn{1}{c}{(K km s$^{-1}$)}  & \multicolumn{1}{c}{(km s$^{-1}$)} & \multicolumn{1}{c}{(km s$^{-1}$)} & \multicolumn{1}{c}{(K)}      &      \\
\hline
\endhead
\hline
\endfoot
\hline
\endlastfoot
215814.58 & 
t-CH$_2$CHCHO $25_{1,25}-24_{1,24}$ & 215814.794(  4) & 137.1 & 24.95 & 5.33 10$^{-4}$ & 0.442( 94) & 6.916( 88) & 0.841(207) & 0.494( 98) & \\ 

215822.35 & 
$^{13}$CH$_3$OCOH $18_{14,5}-17_{14,4}\,E$ & 215822.298( 13) & 227.8 & 7.08 & 5.95 10$^{-5}$ & 0.149( 45) & 6.517( 77) & 0.477(151) & 0.294( 83) & \\

215829.57 &        Unidentified        &   ...   &   ...   &   ...   &   ...   & 0.286( 52) & 6.626( 55) & 0.533(103) & 0.505(110) \\

215832.97 & 
CH$_3$OCOH $\nu_t=2$, $19_{1,18}-18_{1,17}\,A$ & 215833.102(  3) & 441.4 & 52.20 & 1.57 10$^{-4}$ & 0.696( 21) & 6.789(  9) & 0.627( 22) & 1.044( 86) & \\ 

215835.13 &
CH$_3$OCOH $\nu_t=1$, $20_{0,20}-19_{1,19}\,A$ & 215835.178(100) & 299.3 & 18.42 & 2.43 10$^{-5}$ & 0.769( 19) & 6.675( 10) & 0.859( 27) & 0.841(107) & \\

215837.52 &
CH$_3$OCOH $\nu_t=1$, $20_{1,20}-19_{1,19}\,A$ & 215837.587(100) & 299.3 & 19.84 & 1.50 10$^{-4}$ & 1.470(192) & 6.683( 53) & 0.847(136) & 1.631( 61) & \\ 

215839.38 &
CH$_3$OCOH $\nu_t=1$, $20_{0,20}-19_{0,19}\,A$ & 215839.544(100) & 299.3 & 19.84 & 1.50 10$^{-4}$ & 1.088(241) & 6.847( 74) & 0.675(169) & 1.513( 49) & \\ 

215841.90 &
CH$_3$OCOD $18_{-10,8}-17_{-10,7}\,E$ & 215841.936(  6) & 158.1 & 12.45 & 1.06 10$^{-4}$ &    ...     &     ...    &     ...    &    ...     & Blen \\ 
          & 
CH$_3$OCOH $\nu_t=1$, $20_{1,20}-19_{0,19}\,A$ & 215842.029(100) & 299.3 & 18.42 & 2.43 10$^{-5}$ &     ...    &     ...    &    ...     &    ...     & Blen \\ 

215843.47 &
CH$_2$DOH $11_{6,6}-12_{5,8}\,e_1$   & 215843.384(  9) & 293.2 & 1.17 & 3.83 10$^{-6}$ & 1.143(209) & 6.413( 83) & 0.963(219) & 1.114(110) & \\ 
          &
CH$_2$DOH $11_{6,5}-12_{5,7}\,e_1$   & 215843.397(  9) & 293.2 & 1.17 & 3.83 10$^{-6}$ &   ...      &    ...     &     ...    &    ...     & \\ 

215848.10 &        Unidentified        &   ...   &   ...   &   ...   &   ...   & 0.733( 90) & 6.602( 63) & 1.061(153) & 0.649(129) & Ori \\ 

215849.09 &        Unidentified        &   ...   &   ...   &   ...   &   ...   & 0.259(105) & 6.598(125) & 0.605(278) & 0.403(123) & Ori \\

215850.33 &        Unidentified        &   ...   &   ...   &   ...   &   ...   & 0.320(109) & 6.600( 90) & 0.510(188) & 0.589(124) & Ori \\  

215853.70 &
CH$_3$OCOD $18_{10,9}-17_{10,8}\,A$   & 215853.832(  6) & 158.2 & 12.46 & 1.06 10$^{-4}$ & 0.558(173) & 6.699(111) & 0.708(263) & 0.740(105) & \\ 
          &
CH$_3$OCOD $18_{10,8}-17_{10,7}\,A$   & 215853.838(  6) & 158.2 & 12.46 & 1.06 10$^{-4}$ &    ...     &    ...     &     ...    &    ...     & \\ 

215858.22 &
CH$_3$CHO $\nu_t=1$, $11_{3,9}-11_{2,10}\,A$ & 215858.298( 27) & 285.7 &  3.89 & 3.15 10$^{-5}$ & 0.863( 63) & 6.738( 28) & 0.785( 63) & 1.034(108) & \\

215861.71 &
CH$_3$OCOD $18_{10,9}-17_{10,8}\,E$   & 215861.789(  7) & 158.1 & 12.45 & 1.06 10$^{-4}$ & 0.316(201) & 6.685(195) & 0.618(486) & 0.480(153) & \\ 

215863.29 &        Unidentified        &   ...   &   ...   &   ...   &   ...   & 0.408(167) & 6.600(149) & 0.803(410) & 0.478(140) & \\

215864.38 & CH$_2$OHCHO $33_{6,28}-33_{5,29}$ & 215864.490( 10) & 335.8 & 14.66 & 1.39 10$^{-4}$ & 0.303(101) & 6.749( 92) & 0.601(259) & 0.474(139) & \\

215865.68 &
CH$_3$CHO $\nu_t=1$, $11_{1,10}-10_{1,9}\,A$ & 215865.742(  7) & 271.0 & 11.85 & 3.54 10$^{-4}$ & 1.886(210) & 6.653( 50) & 1.180(173) & 1.501(170) & \\ 
\hline
216106.83 &        Unidentified        &   ...   &   ...   &   ...   &   ...   & 0.945(107) & 6.599( 44) & 0.801(109) & 1.109(176) & Ori \\ 

216107.25 &        Unidentified        &   ...   &   ...   &   ...   &   ...   & 0.395(139) & 6.600(109) & 0.621(258) & 0.598(172) \\ 

216109.67 &
CH$_3$OCOH $19_{2,18}-18_{2,17}\,E$   & 216109.735(  6) & 108.7 & 49.37 & 1.49 10$^{-4}$ & 1.873(106) & 6.676( 28) & 1.047( 71) & 1.680(141) & \\ 

216112.60 & DCO$^+$ $3-2$             & 216112.580(  1) &  20.7 &  3.00 & 7.61 10$^{-4}$ & ... & ... & ... & ... & CP \\

216114.63 &
CH$_3$OCOH $29_{9,20}-29_{8,21}\,A$   & 216114.934( 10) & 312.0 & 16.62 & 1.53 10$^{-5}$ & 0.563(  6) & 6.922( 45) & 0.941( 42) & 0.562(168) & \\ 

216115.52 & 
CH$_3$OCOH $19_{2,18}-18_{2,17}\,A$   & 216115.554(  4) & 109.3 & 18.55 & 1.48 10$^{-4}$ & 2.106(109) & 6.727( 28) & 1.165( 76) & 1.699(203) & \\ 

216118.46 &        Unidentified        &   ...   &   ...   &   ...   &   ...   & 0.940( 80) & 6.600( 31) & 0.724( 68) & 1.220(135) & \\

216126.86 &        Unidentified        &   ...   &   ...   &   ...   &   ...   & 0.593(115) & 6.612(113) & 1.249(286) & 0.446(138) & Ori \\

216129.55 &
CH$_2$DOH $12_{0,12}-11_{1,11}\,e_0$ & 216129.626( 10) & 166.5 & 2.25 & 4.21 10$^{-5}$ & 0.833( 55) & 6.457( 49) & 0.732( 67) & 1.069(116) & Ext \\ 
          &                          &                 &       &      &                 & 1.067(183) & 7.143( 81) & 1.097(216) & 0.914 \\

216133.60 &        Unidentified        &   ...   &   ...   &   ...   &   ...   & 0.370(219) & 6.600(234) & 0.752(529) & 0.462(124) & \\

216134.39 &
CH$_3$OCOH $29_{-9,20}-29_{-8,21}\,E$ & 216134.545( 13) & 311.4 & 7.63 & 1.52 10$^{-5}$ & 0.957( 91) & 6.751( 43) & 1.014(125) & 0.887(125) & \\ 

216136.52 &        Unidentified        &   ...   &   ...   &   ...   &   ...   & 0.593(119) & 6.600( 96) & 1.036(251) & 0.538(106) & \\
\hline
217077.08 &
CH$_3$OCOH $30_{4,26}-30_{4,27}\,A$   & 217077.123( 14) & 291.5 & 0.99 & 5.13 10$^{-6}$ & 0.770( 53) & 6.661( 43) & 1.264(101) & 0.572( 66) & \\ 

217084.55 & 
$^{13}$CH$_3$OCOH $18_{-8,10}-17_{-8,9}\,E$ & 217084.731( 13) & 140.9 & 14.40 & 1.23 10$^{-4}$ & 0.274( 47) & 6.761( 60) & 0.668(117) & 0.386( 99) & \\ 

217100.32 & 
$^{13}$CH$_3$OCOH $18_{8,11}-17_{8,10}\,A$  & 217100.401(  4) & 140.9 & 14.45 & 1.24 10$^{-4}$ & 0.476(136) & 6.710(100) & 0.831(307) & 0.538( 79) & \\ 

217102.00 & 
$^{13}$CH$_3$OCOH $18_{8,10}-17_{8,9}\,A$   & 217102.066(  4) & 140.9 & 14.45 & 1.24 10$^{-4}$ & 0.126( 79) & 6.576(115) & 0.367(285) & 0.324(100) & \\

217104.27 & 
$^{13}$CH$_3$OCOH $18_{8,11}-17_{8,10}\,E$  & 217104.070( 13) & 140.9 & 14.40 & 1.23 10$^{-4}$ & 0.181(130) & 6.342(142) & 0.465(490) & 0.366(112) & \\ 

217109.40 &        Unidentified        &   ...   &   ...   &   ...   &   ...   & 0.367( 84) & 6.600( 43) & 0.389(108) & 0.886(144) & \\

217114.74 &        Unidentified        &   ...   &   ...   &   ...   &   ...   & 0.242( 58) & 6.601( 45) & 0.376(110) & 0.606(118) & \\

217117.91 &        Unidentified        &   ...   &   ...   &   ...   &   ...   & 0.858( 69) & 6.600( 34) & 0.871( 84) & 0.926(125) & Ext \\

217130.02 &        Unidentified        &   ...   &   ...   &   ...   &   ...   & 0.574( 88) & 6.602( 72) & 0.913(150) & 0.590(137) & \\

217132.81 &
CH$_3$OD $9_1-8_2\,E$               &  217132.735(390)& 107.3 &   2.80  &               & 1.111(286) & 6.429(155) & 1.203(344) & 0.868( 95) & Ext \\ 
          &                         &                 &       &        &               & 0.103(191) & 7.280(237) & 0.376(504) & 0.258 \\
\hline
217212.15 &        Unidentified        &   ...   &   ...   &   ...   &   ...   & 0.409( 72) & 6.612( 57) & 0.798(175) & 0.481(108) & \\

217212.60 &        Unidentified        &   ...   &   ...   &   ...   &   ...   & 0.708( 90) & 6.602( 48) & 1.216(200) & 0.547(138) & \\

217214.68 &        Unidentified        &   ...   &   ...   &   ...   &   ...   & 0.473( 98) & 6.659( 86) & 1.313(311) & 0.339(149) & \\

217215.68 &
CH$_3$OCOH $32_{9,24}-32_{8,25}\,A$   & 217215.822( 10) & 367.8 & 19.01 & 1.61 10$^{-5}$ & 0.350(116) & 6.841( 85) & 0.581(265) & 0.567(145) & \\ 

217216.28 &        Unidentified        &   ...   &   ...   &   ...   &   ...   & 0.286( 70) & 6.559( 61) & 0.513(151) & 0.524(140) & \\

217225.26 &        Unidentified        &   ...   &   ...   &   ...   &   ...   & 0.405( 90) & 6.644( 68) & 0.603(142) & 0.631(154) & \\

217231.81 &        Unidentified        &   ...   &   ...   &   ...   &   ...   & 0.304(108) & 6.601(135) & 0.778(316) & 0.368(122) & \\

217235.90 &
CH$_3$OCOH $32_{9,24}-32_{8,25}\,E$   & 217235.925( 16) & 367.2 & 8.71 & 1.60 10$^{-5}$ & 0.312(173) & 6.633(174) & 0.623(391) & 0.471(176) & \\ 

217236.92 &  DCN $3-2,\,F=3-3$        & 217236.998(  2) &  20.9 & 0.33 & 5.11 10$^{-5}$ & ... & ... & ... & ... & CP \\

217237.05 & 
CH$_3$OCH$_3$ $34_{7,28}-33_{8,25}\,EA$ & 217236.447(961) & 611.5 &  7.72 & 1.13 10$^{-5}$ &    ...     &    ...     &    ...     &    ...     & Blen \\ 
          & 
CH$_3$OCH$_3$ $34_{7,28}-33_{8,25}\,AE$ & 217236.713(961) & 611.5 & 11.70 & 1.14 10$^{-5}$ &    ...     &    ...     &    ...     &    ...     & Blen \\ 
          & 
CH$_3$OCH$_3$ $34_{7,28}-33_{8,25}\,EE$ & 217237.046(960) & 611.5 & 31.10 & 1.14 10$^{-5}$ &    ...     &    ...     &    ...     &    ...     & Blen \\ 
          & 
CH$_3$OCH$_3$ $34_{7,28}-33_{8,25}\,AA$ & 217237.512(960) & 611.5 & 19.50 & 1.14 10$^{-5}$ &    ...     &    ...     &    ...     &    ...     & Blen \\ 

217238.58 & DCN $3-2$                   & 217238.538(  1) &  20.9 & 3.00  & 4.60 1'$^{-4}$ & ... & ... & ... & ... & CP \\

217240.52 & DCN  $3-2,\,F=2-2$          & 217240.626(  3) &  20.9 & 0.33  & 7.16 10$^{-5}$ & ... & ... & ... & ... & CP \\

217243.88 &        Unidentified        &   ...   &   ...   &   ...   &   ...   & 0.222( 71) & 6.598( 83) & 0.497(172) & 0.420(112) & \\

217248.70 &        Unidentified        &   ...   &   ...   &   ...   &   ...   & 0.484( 73) & 6.585( 56) & 0.724(120) & 0.628(125) & Ori \\

217262.19 &        Unidentified        &   ...   &   ...   &   ...   &   ...   & 0.869(122) & 6.577( 94) & 1.592(209) & 0.513(128) & \\

217262.22 &        Unidentified        &   ...   &   ...   &   ...   &   ...   & 0.546(203) & 6.600( 99) & 0.646(322) & 0.793(120) & \\

217262.67 &
CH$_3$OCOH $37_{10,27}-37_{9,28}\,A$  & 217262.828( 18) & 485.0 & 22.89 & 1.68 10$^{-5}$ & 0.416(187) & 6.920(121) & 0.579(335) & 0.676(124) & \\ 

217266.27 &
CH$_2$DOH  $26_{4,22}-26_{3,24}\,e0$  & 217266.377( 14) & 817.2 & 10.99 & 1.99 10$^{-5}$ & ... & ... & ... & ... & Blen \\    
          &
CH$_3$OCOH $30_{4,26}-30_{3,27}\,A$   & 217266.537( 14) & 291.5 & 12.02 & 1.09 10$^{-5}$ & 0.692(233) & 6.943(135) & 0.873(379) & 0.745(133) & \\
\hline
218439.95 &
CH$_3$OH $4_2-3_1\,E$               & 218440.063( 13) &  37.6 & 1.74 & 4.69 10$^{-5}$ & 1.112(116) & 6.239(464) & 0.803( 28) & 1.301(138) & Ext \\
          &                         &                 &       &      &                 & 0.722( 75) & 7.017( 79) & 0.684(156) & 0.991 \\

218444.87 &
CH$_3$OCOD $20_{-1,19}-19_{2,18}\,E$  & 218444.812(  7) & 115.9 & 3.22 & 2.57 10$^{-5}$ & 0.242( 64) & 6.586( 31) & 0.251(131) & 0.909(177) & \\ 

218448.46 &        Unidentified        &   ...   &   ...   &   ...   &   ...   & 0.369(118) & 6.601(121) & 0.690(189) & 0.502(176) & \\

218449.39 &
CH$_3$OCOD $19_{-2,17}-18_{-2,16}\,E$ & 218449.497(  7) & 112.6 & 18.26 & 1.53 10$^{-4}$ & 1.095(142) & 6.682( 56) & 0.951(158) & 1.082(167) & \\ 

218450.15 &
CH$_3$OCOD $20_{1,19}-19_{2,18}\,A$   & 218450.031(  8) & 115.9 & 15.57 & 2.58 10$^{-5}$ & 0.248( 89) & 6.399( 38) & 0.284(134) & 0.822(166) & \\ 

218451.32 &
CH$_2$CHCN $23_{5,19}-22_{5,18}$  & 218451.297(  1) & 179.8 & 21.91 & 8.26 10$^{-4}$ & 0.623(169) & 6.727(111) & 0.801(265) & 0.730(165) \\

218452.13 &
CH$_2$DOCOH $19_{10,10}-18_{10,9}\,s$ & 218452.118(  3) & 172.7 & 13.74 & 1.14 10$^{-4}$ &    ...     &    ...     &    ...     &    ...     & Blen \\ 
          &
CH$_2$DOCOH $19_{10,9}-18_{10,8}\,s$  & 218452.119(  3) & 172.7 & 13.74 & 1.14 10$^{-4}$ &    ...     &    ...     &    ...     &    ...     & Blen \\ 
          &
CH$_2$CHCN $23_{5,18}-22_{5,17}$  & 218452.356(  1) & 179.8 & 21.91 & 8.26 10$^{-4}$ &    ...     &    ...     &    ...     &    ...     & Blen \\ 

218457.51 &
CH$_3$OCOD $19_{2,17}-18_{2,16}\,A$   & 218457.703(  5) & 112.6 & 18.26 & 1.53 10$^{-4}$ & 0.589(171) & 6.832( 60) & 0.447(167) & 1.238(180) & \\ 

218459.44 &
NH$_2$CHO $10_{1,9}-9_{1,8}$  & 218459.215(  1) &  60.8 & 9.89 & 7.47 10$^{-4}$ & 0.260(215) & 6.306(265) & 0.600(650) & 0.408(146) \\

218461.25 &        
t-CH$_3$CH$_2$OH $5_{3,2}-4_{2,3}$ & 218461.226( 50) &  23.9 & 2.89 & 6.60 10$^{-5}$ & 0.772(137) & 6.567( 80) & 0.963(217) & 0.754(162) & \\

218462.22 & 
NH$_2$CN $11_{1,11}-10_{1,10}$  & 218461.796( 20) &  77.4 & 32.72 & 1.08 10$^{-3}$ & 0.260(153) & 5.989(233) & 0.811(553) & 0.301(143) \\

218463.37 &
CH$_2$CHCN $23_{9,15}-22_{9,14}$  & 218463.739(  1) & 300.3 & 19.48 & 7.34 10$^{-4}$ & 0.304( 81) & 7.094( 71) & 0.509(150) & 0.562(168) \\ 
          &
CH$_2$CHCN $23_{9,14}-22_{9,13}$  & 218463.739(  1) & 300.3 & 19.48 & 7.34 10$^{-4}$ &    ...     &    ...     &     ...    &    ...     \\  

218468.88 & 
aGg'-(CH$_2$OH)$_2$ $22_{14,8}\,0-21_{14,7}\,1$ & 218468.381(  4) & 220.6 & 90.46 & 1.51 10$^{-4}$ & 0.210( 80) & 5.766(108) & 0.537(244) & 0.368(164) & \\
          &
aGg'-(CH$_2$OH)$_2$ $22_{14,9}\,0-21_{14,8}\,1$ & 218468.381(  4) & 220.6 & 116.32 & 1.51 10$^{-4}$&     ...    &     ...    &    ...     &    ...     & \\    

218472.68 &        Unidentified        &   ...   &   ...   &   ...   &   ...   & 0.884(102) & 6.598( 38) & 0.705(101) & 1.178(178) & \\

218474.06 &        Unidentified        &   ...   &   ...   &   ...   &   ...   & 0.992(113) & 6.587( 44) & 0.811(118) & 1.149(165) & \\

218475.68 & 
H$_2$CO p $3_{2,2}-2_{2,1}$    & 218475.634(  1) &  68.1 & 1.67 & 1.57 10$^{-4}$ & 1.316(162) & 6.475( 53) & 0.764(183) & 1.619(174) & Ext \\
          &                         &                 &       &      &                 & 0.649(150) & 7.169(107) & 0.730(354) & 0.835 \\

218479.44 &
CH$_3$OCOD $20_{2,19}-19_{2,18}\,E$   & 218479.619(  7) & 115.9 & 19.52 & 1.56 10$^{-4}$ & 0.708( 99) & 6.645( 50) & 0.729(118) & 0.912(174) & \\ 

218482.46 &       Unidentified        &   ...   &   ...   &   ...   &   ...   & 1.135(120) & 6.601( 60) & 1.129(134) & 0.944(158) & \\

218484.72 &
CH$_3$OCOD $20_{2,19}-19_{2,18}\,A$   & 218484.769(  8) & 115.9 & 19.52 & 1.56 10$^{-4}$ & 0.671( 92) & 6.513( 39) & 0.697(130) & 0.903(136) & \\ 

218489.38 &
CH$_3$OCH$_3$ $23_{3,21}-23_{2,22}\,EA$ & 218489.791(125) & 263.8 & 15.30  & 3.35 10$^{-5}$ & 0.825( 93) & 7.191( 36) & 0.641( 83) & 1.208(159) \\ 
          &
CH$_3$OCH$_3$ $23_{3,21}-23_{2,22}\,AE$ & 218489.791(125) & 263.8 &  7.67 & 3.36 10$^{-5}$ &    ...     &    ...     &    ...     &    ...     \\  
\hline
219539.59 &
t-CH$_3$CH$_2$OCOH $40_{15,25}-39_{15,24}$ & 219539.465( 30) & 377.8 & 34.37 & 1.79 10$^{-4}$ & 0.652(264) & 6.697(253) & 1.211(613) & 0.506(142) \\
          &
t-CH$_3$CH$_2$OCOH $40_{15,26}-39_{15,25}$ & 219539.465( 30) & 377.8 & 34.37 & 1.79 10$^{-4}$ & ... & ... & ... & ... & \\    

219540.21 & 
aGg'-(CH$_2$OH)$_2$ $22_{2,21}\,1-21_{2,20}\,0$ & 219540.442(  2) & 122.2 & 150.06 & 2.54 10$^{-4}$ & 0.348( 88) & 6.947( 62) & 0.606(201) & 0.539(133) & \\

219540.84 &        Unidentified        &   ...   &   ...   &   ...   &   ...   & 0.256( 72) & 6.522( 65) & 0.423(129) & 0.568(102) & Ori \\

219542.89 &        Unidentified        &   ...   &   ...   &   ...   &   ...   & 0.391( 27) & 6.598( 25) & 0.721( 61) & 0.510(145) & Ori \\

219543.61 &        Unidentified        &   ...   &   ...   &   ...   &   ...   & 0.376( 41) & 6.600( 28) & 0.595( 87) & 0.595(138) & Ori \\

219543.97 &        Unidentified        &   ...   &   ...   &   ...   &   ...   & 0.286(110) & 6.600( 95) & 0.495(218) & 0.544(131) & \\

219547.36 &
HNCO $10_{4,6}-9_{4,5}$  & 219547.082( 30) & 708.7 & 6.94 & 1.04 10$^{-4}$ & 0.715(338) & 6.186(262) & 1.086(596) & 0.618(111) \\
          &
HNCO $10_{4,7}-9_{4,6}$  & 219547.082( 30) & 708.7 & 6.94 & 1.04 10$^{-4}$ &     ...    &    ...     &    ...     &    ...     \\  

219551.19 &
CH$_2$DOH $5_{1,5}-4_{1,4}\,e_1$     & 219551.485( 14) &  48.2 & 0.77 & 5.58 10$^{-6}$ & 0.887(178) & 6.641( 93) & 1.047(277) & 0.796(153) & Ext \\ 
          &                          &                 &       &      &                 & 0.718(173) & 7.172(119) & 1.180(402) & 0.572      & \\ 

219560.20 &  C$^{18}$O $2-1$         & 219560.358(  0) &  15.8 & 2.00 & 6.01 10$^{-7}$ & ... & ... & ... & ... & CP \\

219566.17 &
CH$_3$OCOH $\nu_t=1$ $18_{15,3}-17_{15,2}\,A$ & 219566.245(100) & 438.6 & 5.53 & 2.59 10$^{-4}$ & 0.736(285) & 6.650(156) & 0.935(484) & 0.740(151) & \\
          &
CH$_3$OCOH $\nu_t=1$ $18_{15,4}-17_{15,3}\,A$ & 219566.245(100) & 438.6 & 5.53 & 2.01 10$^{-4}$ &     ...    &    ...     &    ...     &    ...     & \\

219568.38 &
CH$_3$OCOH $\nu_t=1$ $18_{14,4}-17_{14,3}\,A$ & 219568.480(100) & 419.2 & 7.15 & 2.60 10$^{-4}$ & 0.807(170) & 6.686( 76) & 0.740(191) & 1.025(122) & \\
          &
CH$_3$OCOH $\nu_t=1$ $18_{14,5}-17_{14,4}\,A$ & 219568.480(100) & 419.2 & 7.15 & 2.13 10$^{-4}$ &    ...     &     ...    &    ...     &    ...     & \\

219569.43 &
t-CH$_3$CH$_2$OCOH $40_{14,26}-39_{14,25}$ & 219569.637( 29) & 357.0 & 35.10 & 1.83 10$^{-4}$ & 0.352( 68) & 6.879( 62) & 0.636(131) & 0.521(110) \\
          &
t-CH$_3$CH$_2$OCOH $40_{14,27}-39_{14,26}$ & 219569.637( 29) & 357.0 & 35.10 & 1.83 10$^{-4}$ & ... & ... & ... & ... & \\

219571.12 &
CH$_3$OCOH $\nu_t=1$ $18_{16,2}-17_{16,1}\,A$ & 219571.198(100) & 459.4 & 3.80 & 2.49 10$^{-4}$ & 0.855(165) & 6.672(125) & 1.256(246) & 0.640(130) & \\
          &
CH$_3$OCOH $\nu_t=1$ $18_{16,3}-17_{16,2}\,A$ & 219571.198(100) & 459.4 & 3.80 & 1.78 10$^{-4}$ &    ...     &     ...    &    ...     &    ...     &  \\

219578.25 &        Unidentified        &   ...   &   ...   &   ...   &   ...   & 0.418(156) & 6.600(137) & 0.845(406) & 0.465(167) & \\

219579.00 &
CH$_3$OCOH $\nu_t=1$ $18_{17,1}-17_{17,0}\,A$ & 219578.701(100) & 481.5 & 1.96 & 2.14 10$^{-4}$ &     ...     &    ...     &    ...     &    ...    & Blen \\
          &
CH$_3$OCOH $\nu_t=1$ $18_{17,2}-17_{17,1}\,A$ & 219578.701(100) & 481.5 & 1.96 & 1.28 10$^{-4}$ &     ...     &    ...     &    ...     &    ...    & Blen \\
          &
CH$_3$OCOH $28_{9,19}-28_{8,20}\,A$   & 219579.069( 10) & 294.6 & 15.77 & 1.58 10$^{-5}$ &    ...     &    ...     &    ...     &    ...     & Blen \\

219580.38 & 
aGg'-(CH$_2$OH)$_2$ $22_{1,21}\,1-21_{1,20}\,0$ & 219580.671(  2) & 122.2 & 195.11 & 2.57 10$^{-4}$ & 0.244( 45) & 6.958( 44) & 0.428( 78) & 0.535(135) & \\

219581.18 &        Unidentified        &   ...   &   ...   &   ...   &   ...   & 0.320( 90) & 6.601( 97) & 0.622(172) & 0.483(143) & \\

219584.23 &
CH$_3$OCOH $\nu_t=1$ $18_{13,5}-17_{13,4}\,A$ & 219584.383(100) & 401.2 & 8.66 & 2.57 10$^{-4}$ & 1.214(159) & 6.696( 62) & 0.988(157) & 1.154(143) & \\
          &
CH$_3$OCOH $\nu_t=1$ $18_{13,6}-17_{13,5}\,A$ & 219584.383(100) & 401.2 & 8.66 & 2.18 10$^{-4}$ &    ...     &    ...     &    ...     &    ...     & \\

219585.80 &
CH$_3$OCOH $30_{9,22}-30_{8,23}\,A$   & 219585.936( 10) & 330.0 & 17.37 & 1.62 10$^{-5}$ & 0.689(132) & 6.710( 66) & 0.677(148) & 0.956( 90) & \\
\hline
219903.58 &        Unidentified        &   ...   &   ...   &   ...   &   ...   & 0.360(105) & 6.600(111) & 0.879(369) & 0.384(130) & \\

219908.38 &
H$_2^{13}$CO o $3_{1,2}-2_{1,1}$ & 219908.481(  7) &  17.8 & 2.67 & 2.56 10$^{-4}$ & 1.045(155) & 6.289( 76) & 1.011(258) & 0.971(118) & Ext \\
          &                        &                 &       &      &                 & 0.990( 28) & 6.967( 58) & 0.754(132) & 1.234 \\

219914.05 &        Unidentified        &   ...   &   ...   &   ...   &   ...   & 0.361(120) & 6.668( 94) & 0.635(253) & 0.534(118) & \\

219914.75 &
H$_2$C$^{17}$O o $3_{1,2}-2_{1,1}$ & 219914.839( 14) &  17.8 & 2.67 & 2.56 10$^{-4}$ & 0.251( 86) & 6.594( 87) & 0.490(199) & 0.481(138) \\

219916.10 &        Unidentified        &   ...   &   ...   &   ...   &   ...   & 0.401( 90) & 6.615( 69) & 0.641(175) & 0.588(166) & \\

219928.49 &        Unidentified        &   ...   &   ...   &   ...   &   ...   & 0.649(137) & 6.600( 88) & 0.821(207) & 0.743(133) & \\

219943.36 &        Unidentified        &   ...   &   ...   &   ...   &   ...   & 0.355(156) & 6.548(144) & 0.679(378) & 0.491(149) & \\

219943.80 &        Unidentified        &   ...   &   ...   &   ...   &   ...   & 0.330(137) & 6.599( 78) & 0.382(177) & 0.814(137) & \\

219944.51 &        Unidentified        &   ...   &   ...   &   ...   &   ...   & 0.351(170) & 6.600(142) & 0.615(377) & 0.536(137) & \\

219944.64 &        Unidentified        &   ...   &   ...   &   ...   &   ...   & 0.326( 92) & 6.677( 86) & 0.525(166) & 0.583(163) & \\

219949.29 &  SO $5_6-4_5$          & 219949.397(  2) &  35.0 & 5.95 & 1.34 10$^{-4}$  & ... & ... & ... & ... & Abs \\

219951.30 &        Unidentified        &   ...   &   ...   &   ...   &   ...   & 0.262(102) & 6.615(136) & 0.649(228) & 0.379(127) & \\

219953.01 &        Unidentified        &   ...   &   ...   &   ...   &   ...   & 0.970(244) & 6.585(141) & 1.197(333) & 0.761(210) & \\

219954.52 &        Unidentified        &   ...   &   ...   &   ...   &   ...   & 0.831(183) & 6.601( 74) & 0.670(171) & 1.164(226) & \\
\hline
220368.00 &
CH$_3$OCOH $\nu_t=1$ $18_{8,11}-17_{8,10}\,A$ & 220368.333(100) & 331.1 & 14.45 & 2.08 10$^{-4}$ &    ...     &    ...     &    ...     &    ...     & Edge \\

220369.73 &
CH$_3$OCOH $\nu_t=1$ $18_{8,10}-17_{8,9}\,A$  & 220369.877(100) & 331.1 & 14.45 & 2.28 10$^{-4}$ & 1.279(137) & 6.655( 48) & 0.953(131) & 1.261(169) \\

220379.43 &        Unidentified        &   ...   &   ...   &   ...   &   ...   & 0.314( 69) & 6.592( 66) & 0.558(129) & 0.529(128) & \\

220385.00 &        Unidentified        &   ...   &   ...   &   ...   &   ...   & 0.378(107) & 6.604(127) & 0.818(252) & 0.434(175) & \\

220391.45 &        Unidentified        &   ...   &   ...   &   ...   &   ...   & 0.716(125) & 6.595( 71) & 0.894(201) & 0.752(157) & \\

220393.75 &        Unidentified        &   ...   &   ...   &   ...   &   ...   & 0.335(146) & 6.575( 90) & 0.455(328) & 0.691(155) & \\

220398.94 &  $^{13}$CO $2-1$           & 220398.684(  0) &  15.9 & 2.00 & 6.08 10$^{-7}$ & ... & ... & ... & ... & CP \\

220401.50 & CH$_3$OH $10_{-5}-11_{-4}\,E$ & 220401.317( 14) & 243.8 & 0.94 & 1.12 10$^{-5}$ & 0.909(244) & 5.969( 99) & 0.746(229) & 1.145(137) & \\
          &                               &                 &       &      &                & 0.669(114) & 6.723(142) & 0.548( 92) & 1.147 & \\

220403.27 &
aGg'-(CH$_2$OH)$_2$ $22_{7,16}\,0-21_{7,15}\,1$ & 220403.609(  3) & 148.8 & 175.42 & 2.34 10$^{-4}$ & ... & ... & ... & ... & Blen \\

220403.86 &        Unidentified        &   ...   &   ...   &   ...   &   ...   & 1.009(697) & 6.598(285) & 0.962(814) & 0.985(175) & \\

220406.50 &        Unidentified        &   ...   &   ...   &   ...   &   ...   & 0.327(123) & 6.599( 96) & 0.508(207) & 0.605(117) & \\

220408.67 &
CH$_3$OCOH $\nu_t=1$ $18_{4,15}-17_{4,14}\,E$ & 220408.752(100) & 299.1 & 17.09 & 1.53 10$^{-4}$ & 1.579(202) & 6.690( 56) & 0.964(155) & 1.539(199) \\

220413.18 &        Unidentified        &   ...   &   ...   &   ...   &   ...   & 0.415(255) & 6.578(204) & 0.596(365) & 0.654(196) & \\

220414.11 &        Unidentified        &   ...   &   ...   &   ...   &   ...   & 0.756(303) & 6.606(143) & 0.724(346) & 0.981(208) & \\

220416.40 &
CH$_3$OCOH $\nu_t=1$ $18_{3,16}-17_{2,15}\,A$ & 220416.305(100) & 293.4 &  9.47 & 1.65 10$^{-5}$ & 0.841(211) & 6.458(101) & 0.870(283) & 0.908(288) & \\

220418.30 &        Unidentified        &   ...   &   ...   &   ...   &   ...   & 0.458(128) & 6.602( 62) & 0.405(128) & 1.063(240) & \\
\hline
230521.40 &        Unidentified        &   ...   &   ...   &   ...   &   ...   & 0.523(155) & 6.600(102) & 0.691(218) & 0.712(172) & \\

230525.86 &
t-CH$_3$CH$_2$OCOH $42_{15,27}-41_{15,26}$ & 230525.888( 33) & 399.6 & 36.64 & 2.10 10$^{-4}$ & 0.729(182) & 6.634(105) & 0.839(237) & 0.816(168) & \\
          &
t-CH$_3$CH$_2$OCOH $42_{15,28}-41_{15,27}$ & 230525.888( 33) & 399.6 & 36.64 & 2.10 10$^{-4}$ & ... & ... & ... & ... & \\

230527.10 &        Unidentified        &   ...   &   ...   &   ...   &   ...   & 0.682(161) & 6.593( 82) & 0.694(193) & 0.924(137) & \\

230528.94 &        Unidentified        &   ...   &   ...   &   ...   &   ...   & 0.450(152) & 6.597(157) & 0.867(268) & 0.487(170) & \\

230537.87 &  CO $2-1$                  & 230538.000(  0) &  16.6 & 2.00 & 6.92 10$^{-7}$ & ... & ... & ... & ... & CP \\

230549.57 &
CH$_3$OCOH $25_{1,24}-25_{0,25}\,A$   & 230549.413( 47) & 182.2 & 3.01 & 3.89 10$^{-6}$ & 0.328(167) & 6.748(128) & 0.483(275) & 0.638(206) & \\ 

230554.51 &
CH$_3$OCOH $25_{2,24}-25_{1,25}\,A$   & 230554.357( 47) & 182.2 & 3.01 & 3.89 10$^{-6}$ & 0.463(138) & 6.346( 68) & 0.516(201) & 0.843(162) & \\ 

230557.38 &        Unidentified        &   ...   &   ...   &   ...   &   ...   & 0.294( 91) & 6.609( 95) & 0.524(228) & 0.527(128) & \\

230560.34 &
CH$_2$DOH  $27_{1,27}-27_{0,27}\,o1$  & 230560.464( 24) & 824.8 & 19.43 & 3.25 10$^{-5}$ & 0.274(122) & 6.598(135) & 0.536(244) & 0.481(100) & \\

230561.10 &        Unidentified        &   ...   &   ...   &   ...   &   ...   & 0.543(167) & 6.653( 90) & 0.698(293) & 0.731(156) & \\

230561.52 &
t-CH$_3$CH$_2$OCOH $42_{14,28}-41_{14,27}$ & 230561.593( 32) & 378.8 & 37.33 & 2.14 10$^{-4}$ & 0.267( 94) & 6.703( 82) & 0.444(175) & 0.564( 88) \\
          &
t-CH$_3$CH$_2$OCOH $42_{14,29}-41_{14,28}$ & 230561.593( 32) & 378.8 & 37.33 & 2.14 10$^{-4}$ & ... & ... & ... & ... & \\    

230564.72 &        Unidentified        &   ...   &   ...   &   ...   &   ...   & 0.396(120) & 6.614(136) & 0.917(349) & 0.406(129) & \\

230572.07 &
CH$_2$DOCOH $21_{2,20}-20_{2,19}\,1a$ & 230572.126(  6) & 123.5 & 54.41 & 1.81 10$^{-4}$ & 0.853(207) & 6.679( 79) & 0.746(250) & 1.073(172) & \\

230573.47 &
CH$_2$DOCOH $21_{2,20}-20_{2,19}\,0a$ & 230573.502(  6) & 123.4 & 54.41 & 1.81 10$^{-4}$ & 0.940(169) & 6.692( 77) & 1.037(238) & 0.851(171) & \\

230576.36 &        Unidentified        &   ...   &   ...   &   ...   &   ...   & 0.486(130) & 6.557( 78) & 0.659(243) & 0.693(155) & \\

230576.92 & 
aGg'-(CH$_2$OH)$_2$ $24_{3,22}\,0-23_{3,21}\,1$ & 230577.144(  2) & 149.8 & 217.11 & 3.04 10$^{-4}$ & ... & ... & ... & ... & Edge \\    
\hline
231195.05 &
CH$_3$CHO $12_{8,4}-11_{8,3}\,E$   & 231195.154(  6) & 216.3 & 7.18 & 2.43 10$^{-4}$ & 1.558(469) & 6.670(150) & 1.042(380) & 1.404(156) & \\ 

231195.65 &
CH$_3$CHO $12_{9,4}-11_{9,3}\,E$   & 231195.518(  7) & 254.5 & 5.66 & 1.91 10$^{-4}$ & 1.523(619) & 6.500(180) & 1.043(534) & 1.372(155) & \\

231199.25 &
gGg'-(CH$_2$OH)$_2$ $24_{2,23}\,1-23_{2,22}\,0$ & 231199.127(  4) & 142.7 & 131.77 & 7.80 10$^{-4}$ & ... & ... & ... & ... & Blen \\    
          &
CH$_3$OCOH $21_{9,12}-21_{8,13}\,A$   & 231199.324(  9) & 190.3 & 10.34 & 1.60 10$^{-5}$ & ... & ... & ... & ... & Blen \\

231199.98 &
CH$_3$CHO $\nu_t=1$ $12_{8,4}-11_{8,3}\,E$  & 231200.043( 16) & 420.8 &  7.17 & 2.42 10$^{-4}$ &    ...     &   ...      &    ...     &    ...     & Blen \\ 
          &
CH$_3$OCOH $21_{-9,12}-21_{-8,13}\,E$ & 231200.172( 10) & 189.7 & 4.73 & 1.58 10$^{-5}$ & ... & ... & ... & ... & Blen \\ 
          &
CH$_2$OHCHO $28_{10,19}-28_{9,20}$ & 231200.383(  4) & 287.1 & 15.07 & 2.06 10$^{-4}$ & ... & ... & ... & ... & Blen \\

231205.56 &        Unidentified        &   ...   &   ...   &   ...   &   ...   & 0.955(200) & 6.597( 99) & 0.993(255) & 0.903(157) & \\

231212.57 &
gGg'-(CH$_2$OH)$_2$ $24_{1,23}\,1-23_{1,22}\,0$ & 231212.070(  4) & 142.7 & 183.93 & 1.09 10$^{-3}$ & ... & ... & ... & ... & Blen \\
          &
CH$_3$CHO $12_{10,2}-11_{10,1}\,A$ & 231212.628(  8) & 297.3 & 3.95 & 1.34 10$^{-4}$ &    ...     &     ...    &    ...     &    ...     & Blen \\ 
          &
CH$_3$CHO $12_{10,3}-11_{10,2}\,A$ & 231212.628(  8) & 297.3 & 3.95 & 1.34 10$^{-4}$ &    ...     &    ...     &    ...     &    ...     & Blen \\

231217.34 &        Unidentified        &   ...   &   ...   &   ...   &   ...   & 0.425(105) & 6.603( 71) & 0.553(148) & 0.723(202) & \\

231218.64 &        Unidentified        &   ...   &   ...   &   ...   &   ...   & 0.436(170) & 6.597(103) & 0.533(243) & 0.769(194) & \\

231220.54 & $^{13}$CS $5-4$ & 231220.685(  3) &  33.3 &  5.00 & 2.51 10$^{-4}$ & ... & ... & ... & ... & CP \\

231222.98 &        Unidentified        &   ...   &   ...   &   ...   &   ...   & 0.622(112) & 6.576( 65) & 0.715(157) & 0.818(145) & \\

231225.47 &
aGg'-(CH$_2$OH)$_2$ $23_{5,19}\,0-22_{5,18}\,1$ & 231225.015(  3) & 148.8 & 145.80 & 2.76 10$^{-4}$ & ... & ... & ... & ... & Blen \\
          &
CH$_3$CHO $12_{7,5}-11_{7,4}\,E$   & 231225.542(  5) & 182.6 & 8.53 & 2.88 10$^{-4}$ & 2.069(353) & 6.682( 82) & 1.033(209) & 1.882(152) & \\

231226.91 &
CH$_3$CHO $12_{9,3}-11_{9,2}\,A$   & 231226.971(  7) & 254.5 & 5.66 & 1.91 10$^{-4}$ & 2.145(115) & 6.688( 28) & 1.250( 88) & 1.612( 94) & \\
          &
CH$_3$CHO $12_{9,4}-11_{9,3}\,A$   & 231226.971(  7) & 254.5 & 5.66 & 1.91 10$^{-4}$ &     ...    &    ...     &    ...     &   ...      & \\

231227.96 &        Unidentified        &   ...   &   ...   &   ...   &   ...   &  0.336( 75) & 6.599( 91) & 0.774(183) & 0.408(129) & Ori \\

231230.50 &
CH$_3$CHO $12_{8,5}-11_{8,4}\,E$   & 231230.603(  6) & 216.2 & 7.18 & 2.43 10$^{-4}$ &    ...     &    ...     &    ...     &    ...     & Blen \\ 
          &
CH$_3$OCOH $\nu_t=1$ $19_{12,7}-18_{12,6}\,E$ & 231230.677(100) & 396.1 & 11.44 & 1.12 10$^{-4}$ &   ...      &    ...     &    ...     &    ...     & Blen \\ 
          &
gGg'-(CH$_2$OH)$_2$ $22_{12,11}\,0-22_{11,11}\,1$ & 231230.780(  9) & 193.4 & 69.02 & 4.45 10$^{-4}$ &   ...      &    ...     &    ...     &    ...     & Blen \\
          &
gGg'-(CH$_2$OH)$_2$ $22_{12,10}\,0-22_{11,12}\,1$ & 231230.781(  9) & 193.4 & 53.69 & 3.46 10$^{-4}$ &   ...      &    ...     &    ...     &    ...     & Blen \\    

231231.90 &
CH$_3$OCOH $29_{4,26}-29_{3,27}\,E$   & 231232.102( 18) & 263.6 & 4.12 & 1.00 10$^{-5}$ & 1.111(370) & 6.772(188) & 1.392(554) & 0.750(168) & \\ 

231235.09 &
gGg'-(CH$_2$OH)$_2$ $27_{12,15}\,0-27_{11,16}\,0$ & 231234.815( 10) & 254.5 & 115.62 & 5.68 10$^{-4}$ &   ...      &    ...     &    ...     &    ...     & Blen \\
          &
gGg'-(CH$_2$OH)$_2$ $27_{12,16}\,0-27_{11,17}\,0$ & 231234.845( 10) & 254.5 & 89.94 & 4.42 10$^{-4}$ &   ...      &    ...     &    ...     &    ...     & Blen \\    
          &
CH$_3$CHO $12_{8,4}-11_{8,3}\,A$   & 231235.139(  6) & 216.3 & 7.18 & 2.43 10$^{-4}$ & 2.047(283) & 6.669( 69) & 1.122(200) & 1.714(195) & \\
          &
CH$_3$CHO $12_{8,5}-11_{8,4}\,A$   & 231235.139(  6) & 216.3 & 7.18 & 2.43 10$^{-4}$ &     ...    &    ...     &     ...    &    ...     & \\

231236.06 &        Unidentified        &   ...   &   ...   &   ...   &   ...   & 0.948(237) & 6.631( 92) & 0.731(216) & 1.218(183) & Ori \\

231237.92 & 
t-CH$_3$CH$_2$OCOH $42_{8,35}-41_{8,34}$ & 231237.632( 24) & 284.2 & 4048 & 2.35 10$^{-4}$ & 0.244( 94) & 6.237( 92) & 0.449(191) & 0.510(200) & \\

231239.11 &
CH$_3$OCOH $21_{9,13}-21_{8,14}\,A$   & 231239.128(  9) & 190.3 & 10.34 & 1.60 10$^{-5}$ & ... & ... & ... & ... & Blen \\
          &
t-CH$_3$CH$_2$OCOH $42_{8,34}-41_{8,33}$ & 231239.260( 24) & 284.2 & 40.48 & 2.35 10$^{-4}$ & ... & ... & ... & ... & Blen \\

231242.76 &        Unidentified        &   ...   &   ...   &   ...   &   ...   & 0.344(214) & 6.598(184) & 0.574(399) & 0.563(197) & \\

231245.20 &
CH$_3$CHO $12_{7,5}-11_{7,4}\,A$   & 231245.034(  5) & 182.6 & 8.53 & 2.88 10$^{-4}$ &     ...    &     ...    &    ...     &    ...     & Blen \\ 
          &
CH$_3$CHO $12_{7,6}-11_{7,5}\,A$   & 231245.034(  5) & 182.6 & 8.53 & 2.88 10$^{-4}$ &     ...    &    ...     &    ...     &    ...     & Blen \\
          &
CH$_3$OCOH $\nu_t=1$ $19_{4,16}-18_{4,15}\,A$ & 231245.424(100) & 310.6 & 18.07 & 2.09 10$^{-4}$ &   ...      &    ...     &    ...     &     ...    & Blen \\ 

231246.44 &        Unidentified        &   ...   &   ...   &   ...   &   ...   & 0.499(150) & 6.603(123) & 0.755(238) & 0.622(148) & \\

231248.12 &
CH$_3$OCOH $29_{4,26}-29_{-2,27}\,E$  & 231248.356( 18) & 263.6 & 2.17 & 5.29 10$^{-6}$ & 0.316( 71) & 6.897( 45) & 0.334( 75) & 0.887(142) & \\
\hline
231303.03 &        Unidentified        &   ...   &   ...   &   ...   &   ...   & 1.101(300) & 6.548(116) & 0.840(278) & 1.232(369) & \\

231310.48 &
CH$_3$CHO $12_{6,7}-11_{6,6}\,E$   & 231310.498(  5) & 153.2 & 9.69 & 3.28 10$^{-4}$ & 2.620(403) & 6.663( 68) & 0.929(173) & 2.650(316) & \\

231311.54 & 
gGg'-(CH$_2$OH)$_2$ $22_{4,18}\,0-21_{4,17}\,1$ & 231311.932(  5) & 134.1 & 121.25 & 7.83 10$^{-4}$ & 0.322(230) & 7.167(158) & 0.407(337) & 0.744(283) & \\

231314.80 &
CH$_3$OCOH $29_{4,26}-29_{3,27}\,A$   & 231315.037( 19) & 264.2 & 8.99 & 1.02 10$^{-5}$ & 1.200(299) & 6.841(127) & 1.005(278) & 1.122(281) & \\ 

231316.70 &        Unidentified        &   ...   &   ...   &   ...   &   ...   & 0.718(201) & 6.580(126) & 0.820(215) & 0.823(256) & \\

231321.86 & N$_2$D$^+$ $3-2$       & 231321.870(  2) &  22.2 & 3.00 & 7.14 10$^{-4}$ & ... & ... & ... & ... & CP \\

231329.53 &
CH$_3$CHO $12_{5,8}-11_{5,7}\,A$   & 231329.639(  4) & 128.6 & 10.68 & 3.62 10$^{-4}$ & 2.128(359) & 6.764( 59) & 0.752(157) & 2.656(432) & \\ 
          &
CH$_3$CHO $12_{5,7}-11_{5,6}\,A$   & 231329.794(  4) & 128.6 & 10.68 & 3.62 10$^{-4}$ & 1.013(196) & 6.408(122) & 0.708(291) & 1.344(420) & \\ 

231331.32 &
CH$_3$OCOH $29_{4,26}-29_{2,27}\,A$   & 231331.224( 19) & 264.2 & 0.81 & 5.26 10$^{-6}$ & 0.477(232) & 6.763(143) & 0.693(459) & 0.647(374) & \\

231346.20 &        Unidentified        &   ...   &   ...   &   ...   &   ...   & 1.151(125) & 6.601( 37) & 0.665( 85) & 1.626(256) & Ori \\

231348.42 &        Unidentified        &   ...   &   ...   &   ...   &   ...   & 1.142(136) & 6.605( 27) & 0.491( 76) & 2.184(270) & Ori \\ 

231350.54 &        Unidentified        &   ...   &   ...   &   ...   &   ...   & 0.985(392) & 6.619(117) & 0.588(282) & 1.573(285) & Ori \\
\hline
231382.04 &
CH$_3$CHO $\nu_t=1$ $12_{5,8}-11_{5,7}\,E$  & 231382.106(  9) & 333.6 & 10.68 & 3.62 10$^{-4}$ & 1.715(379) & 6.733(102) & 0.966(270) & 1.668(100) & Edge \\ 

231386.89 &        Unidentified        &   ...   &   ...   &   ...   &   ...   & 1.765(222) & 6.601( 87) & 1.465(232) & 1.132(141) & Ori \\ 

231388.70 &
$^{13}$CH$_3$OH $21_{-4,18}-20_{-5,15}\,E$ &  231389.062(132) & 603.3 &  6.30 & 1.71 10$^{-5}$ & 0.671(228) & 6.591(146) & 0.896(370) & 0.703(197) & \\ 

231392.11 &        Unidentified        &   ...   &   ...   &   ...   &   ...   & 0.558(102) & 6.608( 42) & 0.531(146) & 0.987(153) & Ori \\ 

231401.90 &
CH$_3$OCOH $40_{-7,33}-40_{-6,34}\,E$ & 231402.296( 31) & 531.1 & 9.78 & 1.74 10$^{-5}$ & 0.292(101) & 7.126(126) & 0.727(298) & 0.378( 93) & \\ 

231409.14 & 
CH$_3$OCOD $16_{4,13}-15_{-3,12}\,E$ & 231408.750( 10) &  88.1 & 1.08 & 1.27 10$^{-5}$ & 0.450(142) & 6.040(135) & 0.744(265) & 0.568(154) & \\

231410.20 &
D$_2$CO o $4_{0,4}-3_{0,3}$        & 231410.224(  2) &  27.9 & 3.99 & 3.47 10$^{-4}$ & 0.442(120) & 5.237(200) & 0.871(200) & 0.477(119) & Ext \\
             &                      &                 &       &      &                 & 0.622(120) & 5.936(200) & 0.705(200) & 0.829 \\
              &                     &                 &       &      &                 & 0.764(120) & 6.572(200) & 0.478(200) & 1.502 \\

231414.41 &
CH$_3$OCOH $35_{10,25}-35_{9,26}\,A$  & 231414.414( 15) & 441.0 & 20.83 & 1.96 10$^{-5}$ &     ...    &    ...     &    ...     &    ...     & Blen \\ 
          &
CH$_3$OCOD $16_{4,13}-15_{3,12}\,A$   & 231414.684( 10) &  88.1 & 5.22 & 1.28 10$^{-5}$ &    ...     &   ...      &    ...     &    ...     & Blen \\ 

231417.13 &
CH$_3$CDO $12_{1,11}-11_{1,10}\,A$ & 231417.149(100) &  74.5 & 26.12 & 8.64 10$^{-4}$ & 1.248(140) & 6.725( 54) & 0.984(131) & 1.192(165) & \\ 

231418.29 &
CH$_3$OCOH $\nu_t=1$ $19_{16,4}-18_{16,3}\,E$ & 231418.457(100) & 470.8 &  5.57 & 5.46 10$^{-5}$ & 0.446( 93) & 6.802( 68) & 0.680(177) & 0.617(163) & \\

231419.61 &
gGg'-(CH$_2$OH)$_2$ $20_{12,8}\,0-20_{11,10}\,1$ & 231419.669(  6) & 172.4 & 45.66 & 3.24 10$^{-4}$ & 0.260(167) & 6.610(340) & 0.800(508) & 0.306(171) & \\
          &
gGg'-(CH$_2$OH)$_2$ $20_{12,9}\,0-20_{11,9}\,1$  & 231419.669(  6) & 172.4 & 58.70 & 4.16 10$^{-4}$ & ... & ... & ... & ... & \\    

231424.28 &
CH$_3$OCOH $38_{8,31}-38_{7,32}\,E$   & 231424.517( 21) & 484.9 & 9.62 & 1.80 10$^{-5}$ & 0.500( 35) & 6.948( 31) & 0.941( 80) & 0.499(109) & \\

231437.23 &
CH$_3$OCOH $38_{8,31}-38_{7,32}\,A$   & 231437.420( 21) & 485.5 & 21.10 & 1.83 10$^{-5}$ & 0.446( 87) & 6.833( 68) & 0.658(176) & 0.637(110) & \\
\end{longtable}
\tablefoot{
Abs: line observed in absorption. 
Blen: line blended with others. 
CP: line with complex line profile. 
Edge: line at the edge of the spectral window. 
Ext: line with extended emission. 
Ori: unidentified line also detected in Orion-KL. 
}
\end{landscape}
}

\begin{table*}
\caption{Obtained abundances with respect to H$_2$ in B1b-S (this work), and towards the B1b envelope observed with the IRAM 30m. 
The D/H ratio obtained for acetaldehyde and methyl formate for the hot and cold components in B1b-S are also shown.}
\label{table:compare}
\centering
\begin{tabular}{lcccccc}
\hline\hline
Species                       & \multicolumn{3}{c}{$X/H_2$\tablefootmark{a}}      & \multicolumn{2}{c}{$D/H$} \\
	   	         &    hot (200 K)   &   cold  (60 K)   &    env.  (10 K)  &    hot (200 K)   &   cold  (60 K)  \\
\hline
NH$_2$CN                         &  7.1 10$^{-14}$  & 9.1 10$^{-14}$ \\
HNCO                                 &  2.1 10$^{-11}$  &       ---        & 1.2 10$^{-10}$ \\
$^{13}$CH$_3$OH             &  3.6 10$^{-10}$  &       ---        & 8.6 10$^{-11}$ \\
CH$_3$CHO                       &  5.7 10$^{-11}$  &  1.4 10$^{-11}$  & 2.0 10$^{-11}$ \\
CH$_3$CHO $\nu_t=1$      &  2.8 10$^{-11}$  &       ---        & \\
CH$_3$CDO                       &  2.8 10$^{-12}$  &  3.6 10$^{-12}$  &                  &   0.05   &  0.25  \\
NH$_2$CHO                       &  2.8 10$^{-13}$  & 3.6 10$^{-13}$ \\
CH$_3$OCH$_3$               &  7.1 10$^{-10}$  &  9.1 10$^{-10}$  & 3.9 10$^{-11}$ \\
$t$-CH$_3$CH$_2$OH      &  1.4 10$^{-11}$  &  1.8 10$^{-11}$  & \\
CH$_2$CHCN                    &  2.8 10$^{-12}$  &  1.8 10$^{-12}$  & 8.2 10$^{-12}$ \\
$t$-CH$_2$CHCHO           &  2.1 10$^{-12}$  &  1.8 10$^{-12}$  & \\
CH$_2$OHCHO                 &  3.6 10$^{-11}$  & 2.7 10$^{-11}$ \\
CH$_3$OCOH                    &  1.1 10$^{-9}$   &  0.4 10$^{-9}$   & 3.9 10$^{-11}$ \\
CH$_3$OCOH $\nu_t=1$   &  1.4 10$^{-10}$  &       ---        & \\
CH$_3$OCOH $\nu_t=2$   &  5.7 10$^{-11}$  &       ---        & \\
$^{13}$CH$_3$OCOH        &  2.1 10$^{-11}$  &  2.7 10$^{-11}$  & \\
CH$_3$OCOD                    &  2.1 10$^{-11}$  &  4.5 10$^{-11}$  &                  &   0.02   &  0.10  \\
CH$_2$DOCOH                 &  2.8 10$^{-11}$  &  9.1 10$^{-11}$  &                  &   0.03   &  0.20  \\
aGg'-(CH$_2$OH)$_2$      & 1.4 10$^{-11}$   & 0.7 10$^{-11}$ \\
gGg'-(CH$_2$OH)$_2$      & 6.4 10$^{-12}$   & 5.4 10$^{-12}$ \\
$t$-CH$_3$CH$_2$OCOH     &  3.3 10$^{-11}$  &  0.7 10$^{-11}$  & \\
\hline
\end{tabular}
\tablefoot{
First values correspond to the compact and hot component (200 K, 0.35$''$), 
second values to the extended and cold one (60 K, 0.60$''$). For vibrational 
excited transitions and HNCO, only the hot component was used considered. 
Abundances for the envelope are from \citet{marce09}, \citet{cerni12}, and 
Marcelino et al. (in prep.). 
Blank spaces mean the molecule was not detected in the IRAM 30m observations.
\tablefoottext{a}{Using N(H$_2$)=1.4 10$^{25}$ cm$^{-2}$ at 0.35'', 
N(H$_2$)=1.1 10$^{25}$ cm$^{-2}$ at 0.60'' (see text), and 
N(H$_2$)=7.6 10$^{22}$ cm$^{-2}$ for the envelope \citep{daniel13}}
}
\end{table*}

\end{appendix}

\end{document}